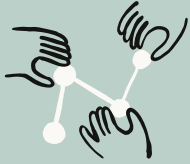

# The Anthropic Economic Index report:
## Uneven geographic and enterprise AI adoption

**Published:**
September 15, 2025

**Authors:**
Ruth Appel*, Peter McCrory*, Alex Tamkin*
Miles McCain, Tyler Neylon, Michael Stern

**Acknowledgements:**

Helpful comments, discussions, and other assistance: Alex Sanchez, Andrew Ho, Ankur Rathi, Asa Kittner, Ben Merkel, Bianca Lindner, Biran Shah, Carl De Torres, Cecilia Callas, Daisy McGregor, Dario Amodei, Deep Ganguli, Dexter Callender III, Esin Durmus, Evan Frondorf, Heather Whitney, Jack Clark, Jakob Kerr, Janel Thamkul, Jared Kaplan, Jared Mueller, Jennifer Martinez, Kaileen Kelly, Kamya Jagadish, Katie Streu, Keir Bradwell, Kelsey Nanan, Kevin Troy, Kim O'Rourke, Kunal Handa, Landon Goldberg, Linsey Fields, Lisa Cohen, Lisa Rager, Maria Gonzalez, Mengyi Xu, Michael Sellitto, Mike Schiraldi, Olivia Chen, Paola Renteria, Rebecca Jacobs, Rebecca Lee, Ronan Davy, Ryan Donegan, Saffron Huang, Sarah Heck, Stuart Ritchie, Sylvie Carr, Tim Belonax, Tina Chin, Zoe Richards

*Lead authors. Contributed equally to this report.

# Introduction

AI differs from prior technologies in its unprecedented adoption speed. In the US alone, 40% of employees report using AI at work, up from 20% in 2023 two years ago.[1] Such rapid adoption reflects how useful this technology already is for a wide range of applications, its deployability on existing digital infrastructure, and its ease of use—by just typing or speaking—without specialized training. Rapid improvement of frontier AI likely reinforces fast adoption along each of these dimensions.

Historically, new technologies took decades to reach widespread adoption. Electricity took over 30 years to reach farm households after urban electrification. The first mass-market personal computer reached early adopters in 1981, but did not reach the majority of homes in the US for another 20 years. Even the rapidly-adopted internet took around five years to hit adoption rates that AI reached in just two years.[2]

Why is this? In short, it takes time for new technologies—even transformative ones—to diffuse throughout the economy, for consumer adoption to become less geographically concentrated, and for firms to restructure business operations to best unlock new technical capabilities. Firm adoption, first for a narrow set of tasks, then for more general purpose applications, is an important way that consequential technologies spread and have transformative economic effects.[3]

**In other words, a hallmark of early technological adoption is that it is *concentrated*—in both a small number of geographic regions and a small number of tasks in firms.** As we document in this report, AI adoption appears to be following a similar pattern in the 21st century, albeit on shorter timelines and with greater intensity than the diffusion of technologies in the 20th century.

To study such patterns of early AI adoption, we extend the [Anthropic Economic Index](#) along two important dimensions, introducing a geographic analysis of Claude.ai conversations and a first-of-its-kind examination of enterprise API use. We show how Claude usage has evolved over time, how adoption patterns differ across regions, and—for the first time—how firms are deploying frontier AI to solve business problems.

## Changing patterns of usage on Claude.ai over time

In the first chapter of this report, we identify notable changes in usage on Claude.ai over the previous eight months, occurring alongside improvements in underlying model capabilities, new product features, and a broadening of the Claude consumer base.

We find:

- **Education and science usage shares are on the rise.** While the use of Claude for coding continues to dominate our total sample at 36%, educational tasks surged from 9.3% to 12.4%, and scientific tasks from 6.3% to 7.2%.

- **Users are entrusting Claude with more autonomy.** "Directive" conversations, where users delegate complete tasks to Claude, jumped from 27% to 39%. We see increased program creation in coding (+4.5pp) and a reduction in debugging (-2.9pp)—suggesting that users might be able to achieve more of their goals in a single exchange.

## The geography of AI adoption

For the first time, we release geographic cuts of Claude.ai usage data across 150+ countries and all U.S. states. To study diffusion patterns, we introduce the Anthropic AI Usage Index (AUI) to measure whether Claude.ai use is over- or underrepresented in an economy relative to its working age population.

We find:

- **The AUI strongly correlates with income across countries.** As with previous technologies, we see that AI usage is geographically concentrated. Singapore and Canada are among the highest countries in terms of usage per capita at 4.6x and 2.9x what would be expected based on their population, respectively. In contrast, emerging economies, including Indonesia at 0.36x, India at 0.27x and Nigeria at 0.2x, use Claude less.

- **In the U.S., local economy factors shape patterns of use.** DC leads per-capita usage (3.82x population share), but Utah is close behind (3.78x). We see evidence that regional usage patterns reflect distinctive features of the local economy: For example, elevated use for IT in California, for financial services in Florida, and for document editing and career assistance in DC.

- **Leading countries have more diverse usage.** Lower-adoption countries tend to see more coding usage, while high-adoption regions show diverse applications across education, science, and business. For example, coding tasks are over half of all usage in India versus roughly a third of all usage globally.
- **High-adoption countries show less automated, more augmented use.** After controlling for task mix by country, low AUI countries are more likely to delegate complete tasks (automation), while high-adoption areas tend toward greater learning and human-AI iteration (augmentation).

The uneven geography of early AI adoption raises important questions about economic convergence. Transformative technologies of the late 19th century and the early 20th centuries—widespread electrification, the internal combustion engine, indoor plumbing—not only ushered in the era of modern economic growth but accompanied a large divergence in living standards around the world.[4]

If the productivity gains are larger for high-adoption economies, current usage patterns suggest that the benefits of AI may concentrate in already-rich regions—possibly increasing global economic inequality and reversing growth convergence seen in recent decades.[5]

## Systematic enterprise deployment of AI

In the final chapter, we present first-of-its-kind insight on a large fraction of our first-party (1P) API traffic, revealing the tasks companies and developers are using Claude to accomplish. Importantly, API users access Claude programmatically, rather than through a web user interface (as with Claude.ai). This shows how early-adopting businesses are deploying frontier AI capabilities.

We find:

- **1P API usage, while similar to Claude.ai use, differs in specialized ways.** Both 1P API usage and Claude.ai usage focus heavily on coding tasks. However, 1P API usage is higher for coding and office/admin tasks, while

Claude.ai usage is higher for educational and writing tasks.

- **1P API usage is automation dominant.** 77% of business uses involve automation usage patterns, compared to about 50% for Claude.ai users. This reflects the programmatic nature of API usage.

- **Capabilities seem to matter more than cost in shaping business deployment.** The most-used tasks in our API data tend to cost more than the less frequent ones. Overall, we find evidence of weak price sensitivity. Model capabilities and the economic value of feasibly automating a given task appears to play a larger role in shaping businesses' usage patterns.

- **Context constrains sophisticated use.** Our analysis suggests that curating the right context for models will be important for high-impact deployments of AI in complex domains. This implies that for some firms costly data modernization and organizational investments to elicit contextual information may be a bottleneck for AI adoption.

## Open source data to catalyze independent research

As with previous reports, we have open-sourced the underlying data to support independent research on the economic effects of AI. This comprehensive dataset includes task-level usage patterns for both Claude.ai and 1P API traffic (mapped to the O*NET taxonomy as well as bottom-up categories), collaboration mode breakdowns by task, and detailed documentation of our methodology. At present, geographic usage patterns are only available for Claude.ai traffic.

**Key questions we hope this data will help others to investigate include:**

- What are the local labor market consequences for workers and firms of AI usage & adoption?

- What determines AI adoption across countries and within the US? What can be done to ensure that the benefits of AI do not only accrue to already-rich economies?

- What role, if any, does cost-per-task play in shaping enterprise deployment patterns?

- Why are firms able to automate some tasks and not others? What implications does this have for which types of workers will experience better or worse employment prospects?

---

1   Gallup 2025, *AI Use at Work Has Nearly Doubled in Two Years*.

2   Bick, Blandin, Deming, 2024 *The Rapid Adoption of Generative AI* benchmark AI adoption against adoption of PC and the internet; Lewis & Severnini, 2020 *Short- and long-run impacts of rural electrification: Evidence from the historical rollout of the U.S. power grid* analyze the impact of bringing electricity to rural areas on economic outcomes.

3   Kalyani, Bloom, Carvalho, Hassan, Lerner and Ahmed Tahoun 2025 *Diffusion of New Technologies*.

4   See Gordon, 2012 *Is U.S. Economic Growth Over? Faltering Innovation Confronts the Six Headwinds* for a comparison of early and late 20th century innovations and their impact on productivity. Pritchett, 1997. *Divergence, Big Time* documents economic divergence that accompanied transition to era of modern economic growth.

5   Kremer, Willis, You, 2022 *Converging to Convergence* present evidence of growth convergence in recent decades. See Jones, Jones, and Aghion, 2017 *Artificial Intelligence and Economic Growth* for discussion of growth implications AI-powered automation of innovation.

Chapter 1:

# Claude.ai Usage Over Time

## Overview

Understanding how AI adoption evolves over time can help predict its economic impacts—from productivity gains to workforce changes. With data spanning from December 2024 and January 2025 (from our first report, 'V1') to February and March 2025 ('V2') to our newest insights from August 2025 ('V3'), we can track how AI usage has shifted over the past eight months as capabilities and product features have improved, new kinds of users have adopted the technology, and uses have become more sophisticated. We view the evidence presented below as suggesting that new product features have enabled new forms of work rather than simply accelerating adoption for existing tasks.

## How Claude.ai usage for economic tasks has changed

### Educational and scientific tasks continue their rise in relative importance

While computer and mathematical tasks still dominate overall usage at 36%, we are seeing sustained growth in knowledge-intensive fields. Educational Instruction and Library tasks rose from 9% in V1 to 12% in V3. Life, Physical, and Social Science tasks increased from 6% to 7%. Meanwhile, the relative share of Business and Financial Operations tasks fell from 6% to 3%, and Management dropped from 5% to 3%.

This divergence suggests AI usage may be diffusing especially quickly among tasks involving knowledge synthesis and explanation, compared to traditional business operations—possibly because these tasks benefit more from Claude's reasoning capabilities.

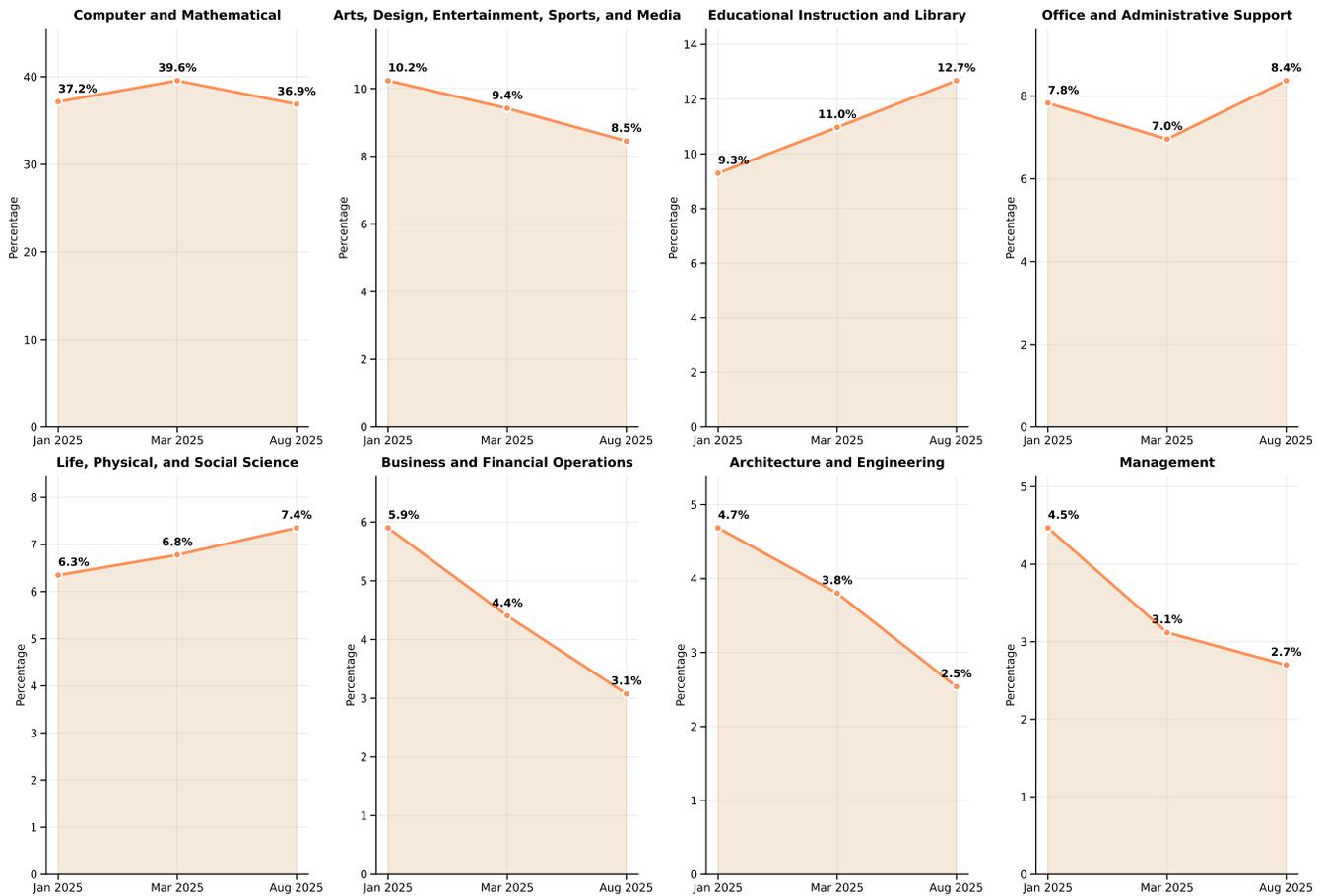

**Figure 1.1: Claude.ai usage over time** Each panel shows the share of sampled conversations on Claude.ai associated with tasks from each SOC major group. We see notable increases in usage for scientific and educational tasks. SOC major groups ranked by usage in our first report.

**New capabilities are shaping usage patterns**

At a more granular level, we document changes in task composition that appear linked to features launched between V2 and V3. For example, searching electronic sources and databases grew substantially (0.03% → 0.49%), likely reflecting our [web search release](#) in March. In addition, we also see a rise in internet-based research tasks (0.003% → 0.27%), which aligns with the [Research](#) mode we released in April.[1]

We also see other kinds of changes. Tasks relating to developing instructional materials increased by 1.3pp, growing from a base of 0.2% to 1.5%—a more than 6-fold increase that may reflect growing [adoption among educators.](#)[2] Creating multimedia documents rose 0.4pp, nearly tripling from 0.16% to 0.55%, potentially driven by continued use of our [Artifacts feature](#) for building traditional and AI-powered apps within Claude.ai.[3]

Interestingly, the share of tasks involving creating new code more than doubled, increasing by 4.5 percentage points (from 4.1% to 8.6%), while debugging and error correction tasks fell by 2.8 percentage points (from 16.1% to 13.3%)—a net 7.4pp shift toward creation over fixing code. This may suggest that models have become increasingly reliable, such that users spend less time fixing problems and more time creating things in a single interaction.

**Directive automation is accelerating**

As in previous reports, we also track not just what people use Claude for but how they collaborate with or delegate to Claude on Claude.ai.

At a high level, we distinguish between *automation* and *augmentation* modes of using Claude:

*Automation* encompasses interaction patterns focused on task completion:
- **Directive:** Users give Claude a task and it completes it with minimal back-and-forth
- **Feedback Loops:** Users automate tasks and provide feedback to Claude as needed

*Augmentation* focuses on collaborative interaction patterns:
- **Learning:** Users ask Claude for information or explanations about various topics
- **Task Iteration:** Users iterate on tasks collaboratively with Claude
- **Validation:** Users ask Claude for feedback on their work

The share of directive conversations sampled from Claude.ai conversations jumped from 27% in V1 in late 2024 to 39% in V3. This increase came primarily at the expense of *task iteration* and *learning* interactions, implying a sizable net increase in the share of conversations exhibiting automative patterns of use – a notable increase in just eight months. This is the first report where automation usage exceeds augmentation usage.

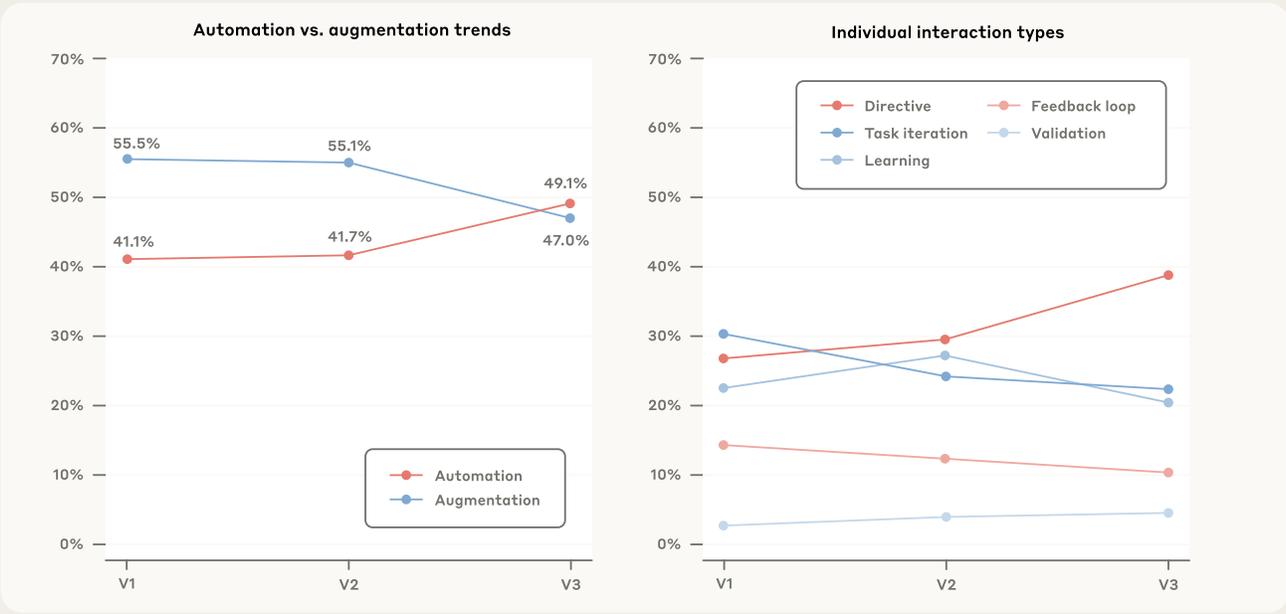

**Figure 1.2: Collaboration mode frequencies across Anthropic Economic Index Reports** The left panel calculates the share of conversations exhibiting either automation or augmentation forms of use. The right panel breaks this out by collaboration mode. Claude tends to be used in more automated ways over time, driven primarily by an increase in directive use.

One interpretation is that this is a result of increasing model capabilities. As models improve at anticipating user needs and producing high-quality outputs on first attempts, users may need fewer follow-up refinements. The jump in directive usage could also signal growing confidence in delegating complete tasks to AI, a form of learning-by-doing.[4]

Whether the growth in directive usage is attributable to improving model capabilities or learning-by-doing could signal very different labor market implications. If more advanced models simply expand the set of automated tasks, then the risk increases that workers performing such tasks will be displaced. However, if instead the rise in directive use reflects learning-by-doing, then workers most able to adapt to new AI-powered workflows are likely to see greater demand and higher wages. In other words, AI may benefit some workers more than others: it may lead to higher wages for those with the greatest ability to adapt to technological change, even as those with lower ability to adapt face job disruption.[5] This will be an important area of inquiry for future research.

**Looking Ahead**

The V3 data reveals that AI capabilities and adoption are continuing to progress. Knowledge-based tasks, including educational and scientific applications, continue their fast growth rate, and new product features appear to be enabling different types of work rather than just accelerating existing tasks.

Most strikingly, the data point toward increased delegation of tasks to AI systems–perhaps due to some combination of user trust in the technology as well as improvement of underlying model capabilities. This could also be due to changes in the underlying user base. The next chapter of this report for the first time breaks down usage across geography, allowing us to disentangle temporal vs. geographic changes more clearly going forward. We will continue to track these trends closely in future reports.

---

1   "Search electronic sources, such as databases or repositories, or manual sources for information" increased from 0.03% to 0.49%. "Conduct internet-based and library research" increased from 0.003% to 0.27%.

2   Statistic computed from tasks containing the string "develop instructional materials".

3   Tasks were collated from the set of tasks whose frequency has changed by a magnitude greater than or equal to 0.2 percentage points. Programming creation tasks include: "write new programs or modify existing programs" (1.5% 4.9%), "design, build, or maintain web sites" (1.2% 2.0%), "write, analyze, review, and rewrite programs" (1.2% 0.5%), "develop new software applications" (0.06% 0.6%), "develop transactional web applications" (0.1% 0.3%), and "develop application-specific software" (0.05% 0.3%). Debugging/error correction tasks include: "modify existing software to correct errors" (two variants: 2.5% 3.8% and 4.8% 2.7%), "correct errors by making appropriate changes" (3.0% 2.1%), "perform initial debugging procedures" (2.0% 0.9%), "diagnose, troubleshoot, and resolve hardware/software problems" (1.6% 2.5%), "review and analyze computer printouts to locate code problems" (1.3% 0.9%), and "determine sources of web page or server problems" (0.9% 0.4%).

4   We note that V3 uses Claude Sonnet 4 for classification, while V2 used Sonnet 3.7, which complicates direct comparison. To address this, we reran V3 data with Sonnet 3.7 and still found directive interactions rising significantly (though to a lower absolute level of 45% automation versus 49% with Sonnet 4). We also verified this trend is not driven by changes in task mix—the shift toward directive interactions appears across a wide range of occupational categories, suggesting it reflects genuine changes in how people interact with Claude rather than compositional effects.

5   Nelson and Phelps, 1966 Investment in Humans, Technological Diffusion, and Economic Growth is a classic reference for the value of education in equipping workers to adapt to change. See also Goldin and Katz, 2008 The Race between Education and Technology. We thank Anton Korinek for the observation that AI itself might accelerate the diffusion and economic impact of AI to the extent that it plays the role that skilled workers played in the past in figuring out how to effectively wield new technologies in novel settings.

Chapter 2:

# Claude usage across the United States and the globe

## Overview

Where AI gets adopted first—and how it's used—will shape economic outcomes across the world. By analyzing Claude usage patterns across 150+ countries and all US states, we uncover three key dynamics: where early adopters are, what they're using AI for, and how usage evolves as adoption matures. These geographic patterns provide real-world evidence about AI's economic diffusion, helping track whether different regions are converging or diverging in their AI adoption, and revealing how local economic characteristics shape technology deployment.

Our data, relying on a [privacy-preserving](privacy-preserving)[1] analysis of 1 million Claude.ai conversations[2], confirmed some of our expectations while challenging others. The US dominates total usage at 21.6%, which is unsurprising given its size and high income. But even when adjusting for the working-age population size, higher-income countries tend to have higher usage. For example, Singapore's usage rate is 4.5 times what its working-age population would suggest, while large regions of the globe show minimal usage. Interestingly, within the US, DC and Utah outpace California in usage per capita.

We also observe changes in AI use cases as adoption per capita deepens. Countries with lower AI adoption per capita concentrate overwhelmingly on coding tasks—over half of all usage in India, compared to roughly a third globally. As adoption matures, usage diversifies, with a rising emphasis on education, science, and business operations.

Even more striking: mature markets tend to use AI more collaboratively, while emerging markets are more likely to delegate complete tasks to it—perhaps reflecting differences in how AI is deployed by economies at different stages of structural transformation. Our data provides a window into these patterns across geographies, and going forward, will enable us to track whether these adoption gaps narrow, widen, or change in structure over time.

## Claude diffusion across the globe

**Total Claude usage is highest in the US**

Claude adoption overall is highly geographically concentrated. In terms of total global usage, the United States accounts for the highest share (21.6%), with the next highest usage countries showing significantly lower shares (India at 7.2%, Brazil at 3.7%, see Figure 2.1). However, this concentration is affected by the population size of each country[3] – larger countries may have larger usage shares purely because of their population size.

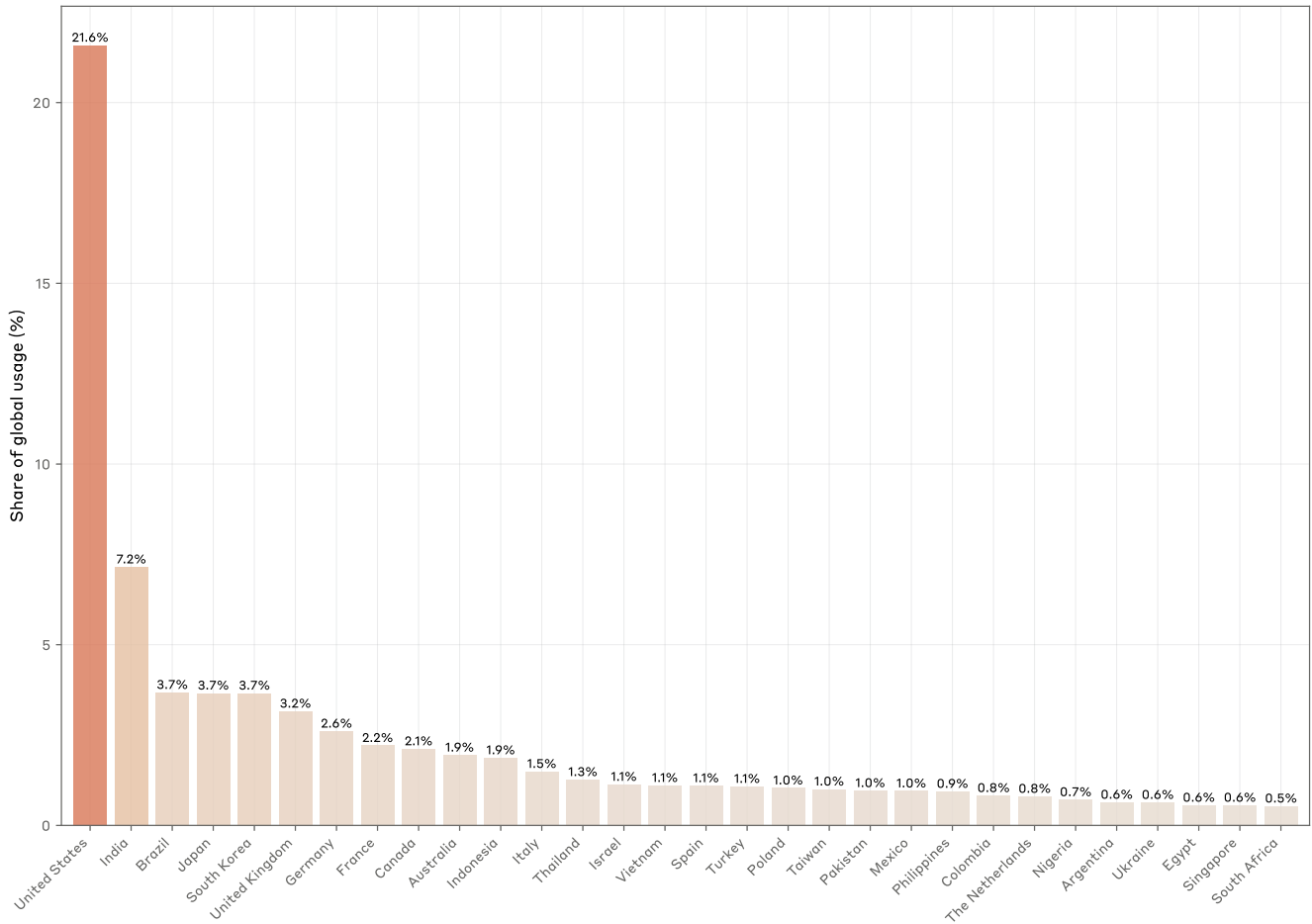

**Figure 2.1: Leading countries in terms of global Claude.ai usage share** The data includes Claude.ai Free and Pro conversations.

**Per capita usage of Claude is concentrated in technologically advanced countries**

To account for differences in population size, we analyze usage adjusted for the working-age population, introducing a new measure called the **Anthropic AI Usage Index (AUI):** For each geography, we calculate its share of Claude usage, and its share of the working-age population (ages 15-64). We then calculate the AUI by dividing these shares:

This index reveals whether countries use Claude more or less than expected relative to their working-age population. A region with an AUI > 1 has higher usage than expected after adjusting for population, while a region with an AUI < 1 has lower usage.

The results reveal a striking pattern of concentration among small, technologically advanced economies. Israel leads global per capita Claude usage with an Anthropic AI Usage Index of 7 — meaning its working-age population uses Claude 7x more than expected based on its population. Singapore follows at 4.57, while Australia (4.10), New Zealand (4.05) and South Korea (3.73) round out the top five countries in terms of per capita Claude usage.

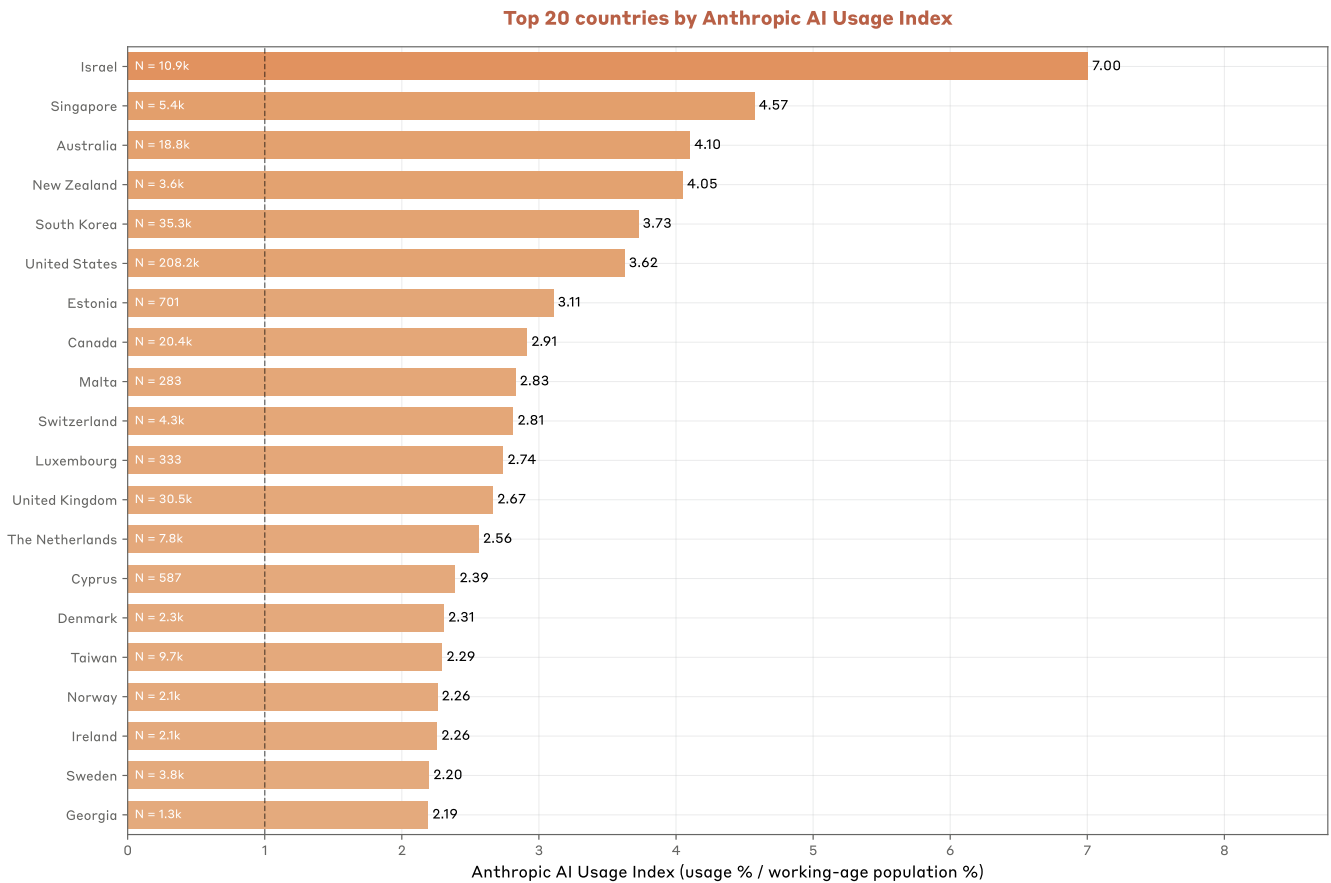

**Figure 2.2: Small, technologically advanced countries are leading in Claude adoption per capita** The figure shows the top 20 countries based on the Anthropic AI Usage Index. We only include countries with at least 200 observations in our sample for this figure because of the uncertainty of the measure for low-usage countries in our random sample. The underlying data includes Claude.ai Free and Pro usage.

Next, we create per capita usage tiers based on the AUI. We look at countries with at least 200 conversations in our random sample of 1 million conversations, and set thresholds for different usage tiers-based quartiles, i.e. *Leading* (top 25%), *Upper Middle* (50-75%), *Lower Middle* (25%-75%) and *Emerging* (bottom 25%). We then assign countries, even if they have fewer than 200 observations, to a tier based on their AUI. We assign countries for which we have population data, but no usage in our sample, to a *Minimal* tier.[4] Figure 2.3 illustrates the Anthropic AI Usage Index tiers across the globe, Table 2.1 shows an overview of the tiers and country examples.

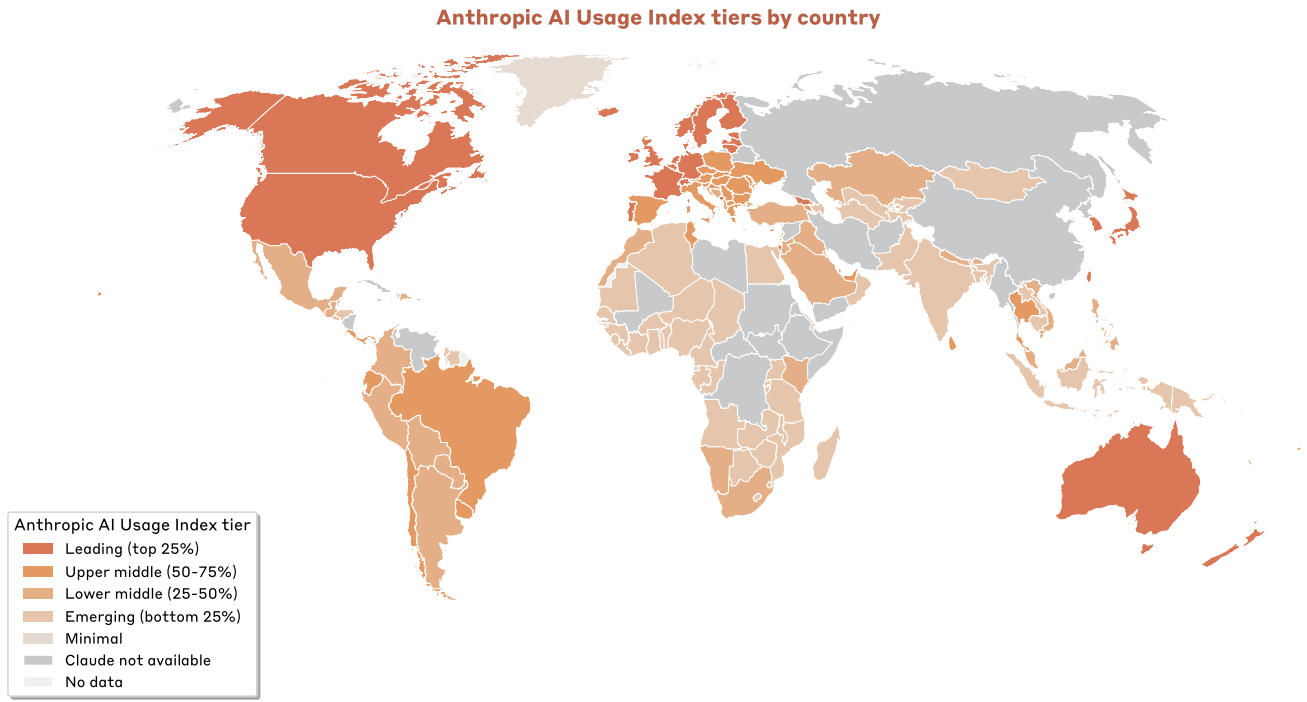

**Figure 2.3: Claude diffusion varies across countries, with countries in North America, Europe and Oceania leading in Claude adoption per working-age capita** The different tiers reflect a country's position within the global distribution of the Anthropic AI Usage Index as defined in this chapter.[5,6]

| Tier | AUI range | # of countries | Example countries |
|---|---|---|---|
| Leading (top 25%) | 1.84 - 7.00 | 37 | Israel, Monaco, Singapore, Australia, New Zealand |
| Upper middle (50-75%) | 0.89 - 1.71 | 35 | Czechia, Austria, Slovenia, Poland, Armenia |
| Lower middle (25-50%) | 0.37 - 0.85 | 39 | Peru, Seychelles, Colombia, Albania, Argentina |
| Emerging (bottom 25%) | 0.01 - 0.36 | 53 | Indonesia, Ghana, Kuwait, Mongolia, Rwanda |
| Minimal | 0.00 - 0.00 | 25 | Aruba, Tonga, Nauru, Samoa, Palau |

**Table 2.1: Anthropic Economic Index tiers with examples, number of countries, and AUI range for each tier.** Zooming into This concentration in advanced economies with limited population sizes reflects their established patterns as technology pioneers. For example, both Israel and Singapore rank highly in the [Global Innovation Index](#)—a measure of how innovative different economies across the globe are—suggesting that general investments in information technology position economies well for rapid adoption of frontier AI. Overall, these economies can leverage their educated workforces, robust digital infrastructure, and innovation-friendly policies to create fertile conditions for AI.

Notable is the position of major developed economies in Claude usage. The United States (3.62) ranks among leading countries in terms of per capita adoption, with Canada (2.91) and the United Kingdom (2.67) having elevated but more moderate rates of adoption as compared to their population. Other major economies show lower adoption, including France at 1.94, Japan at 1.86, and Germany at 1.84.

Meanwhile, many lower and middle-income economies show minimal Claude usage, with many countries across Africa, Latin America, and parts of Asia showing Claude adoption below what would be expected based on their working-age population. This includes Bolivia (0.48), Indonesia (0.36), India (0.27), and Nigeria (0.2).

This variation in usage is reflective of income differences across these economies. We see a strong positive correlation between Claude adoption and Gross Domestic Product per working-age capita (see Figure 2.4), with a 1% increase in GDP per capita being associated with a 0.7% increase in Claude usage per capita.

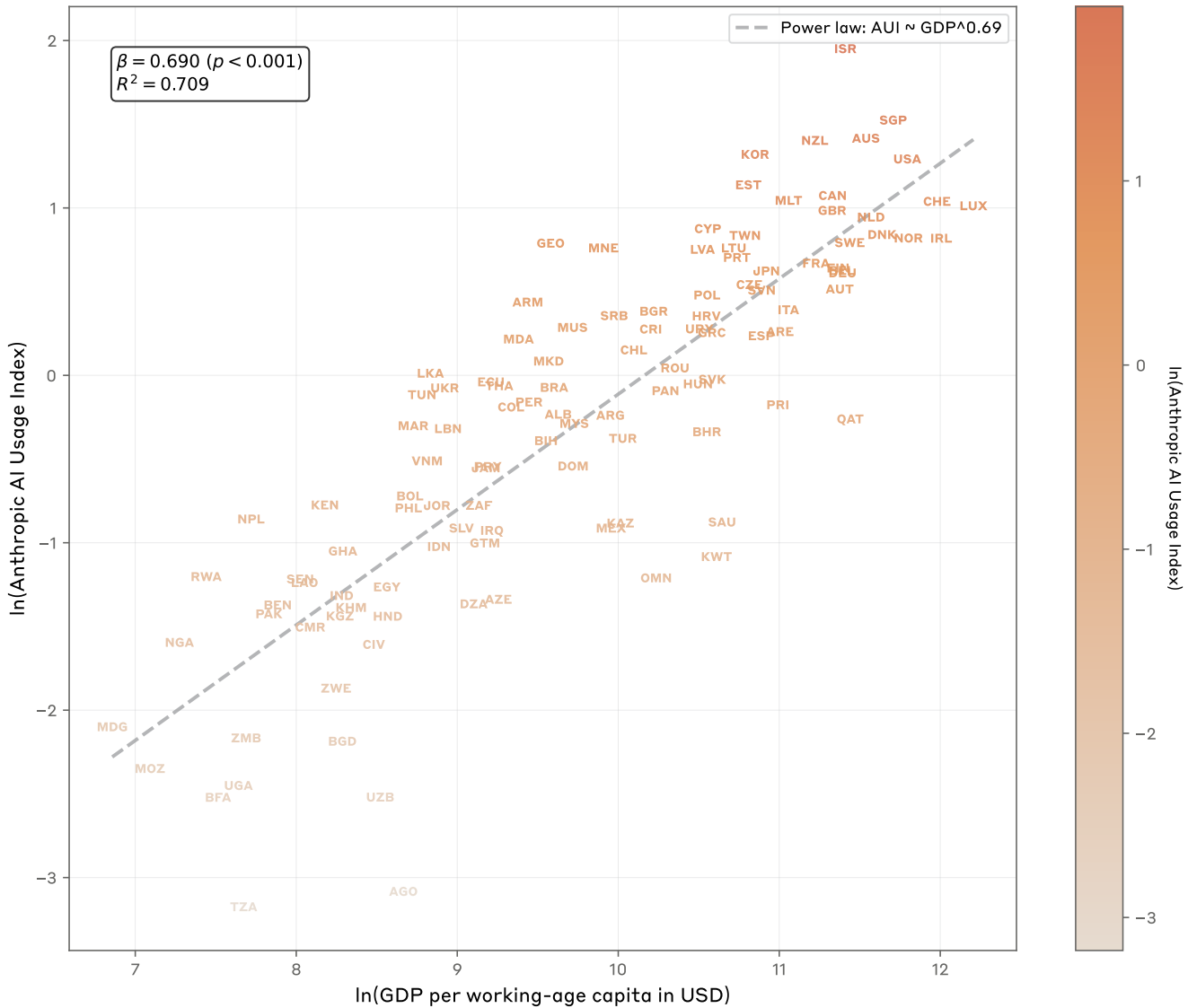

**Figure 2.4: Claude usage per capita is positively correlated with income per capita across countries** We only include countries with at least 200 observations in our sample for this figure because of the uncertainty of the measure for low-usage countries in our random sample. Axes are on a log scale, highlighting a power law distribution. Each country is represented by its 3-letter ISO code.

The disparities in Claude usage likely reflect a confluence of factors, some of which are correlated with income:

- **Digital infrastructure:** High-usage countries typically have robust internet connectivity and cloud computing access needed to access AI assistants.

- **Economic structure:** As documented in this and previous reports, Claude capabilities are well-suited to various tasks typical of knowledge workers. Advanced economies tend to have a greater share of the workforce in such roles as compared to lower-income economies with a larger employment

- share in manufacturing.
- **Regulatory environment:** Governments differ in how actively they encourage the use of AI across different industries and in how heavily they regulate the technology.
- **Awareness and access:** Countries with stronger connections to Silicon Valley and AI research communities may have greater awareness of and access to Claude.
- **Trust and comfort:** Public opinion on trust in AI varies substantially [across countries](#).

## Claude diffusion across the United States

Within the US, California overwhelmingly leads with 25.3% of usage. Other states with major tech centers like New York (9.3%), Texas (6.7%), and Virginia (4.0%) also rank highly. Though not adjusted for population, we suspect that these strong adoption figures partly reflect rapid adoption in technology hubs—in keeping with how economically consequential technologies have historically tended to diffuse.

This narrative becomes more complex, however, when we adjust for the population size of each state. Surprisingly, the District of Columbia leads with an Anthropic AI Usage Index of 3.82, indicating that Claude usage in DC is 3.82x greater than its share of the country's working-age population. Closely following is Utah (3.78), notably ahead of California (2.13), New York (1.58) and Virginia (1.57).[7]

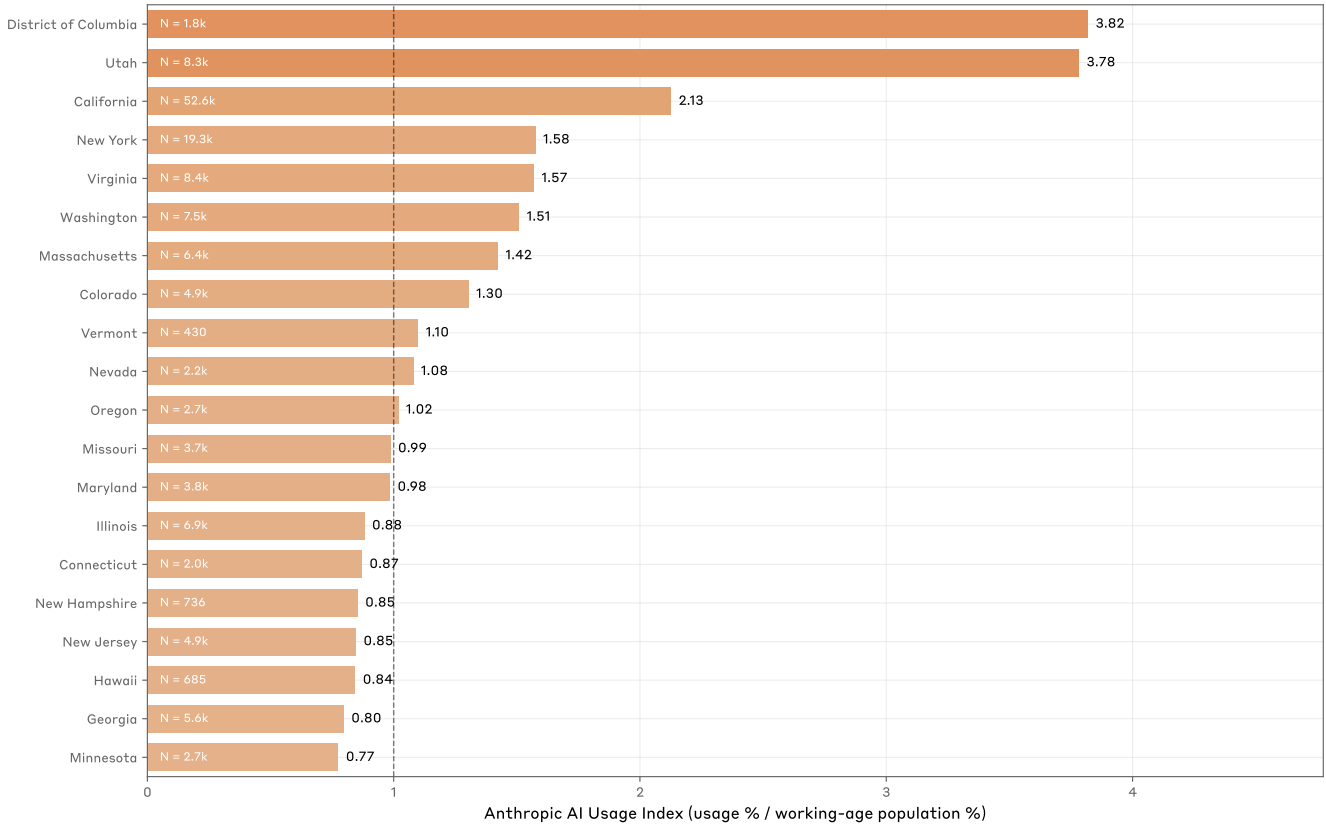

**Figure 2.5: Leading US states in terms of Claude adoption per working-age capita include the District of Columbia, Utah, California, New York and Virginia** The figure shows the top 20 US states based on the Anthropic AI Usage Index. We only include states with at least 100 observations in our sample for this figure because of the uncertainty of the measure for low-usage states in our random sample. The underlying data includes Claude.ai Free and Pro usage.

We document a similar, but weaker correlation than at the global level between Claude adoption and income per capita across US states. Income differences explain less than half the variation in cross-state adoption rates. Despite this weaker correlation, we find that Claude adoption rises faster with income: Each 1% increase in state GDP per capita is associated with a 1.8% increase in the AI Usage Index.

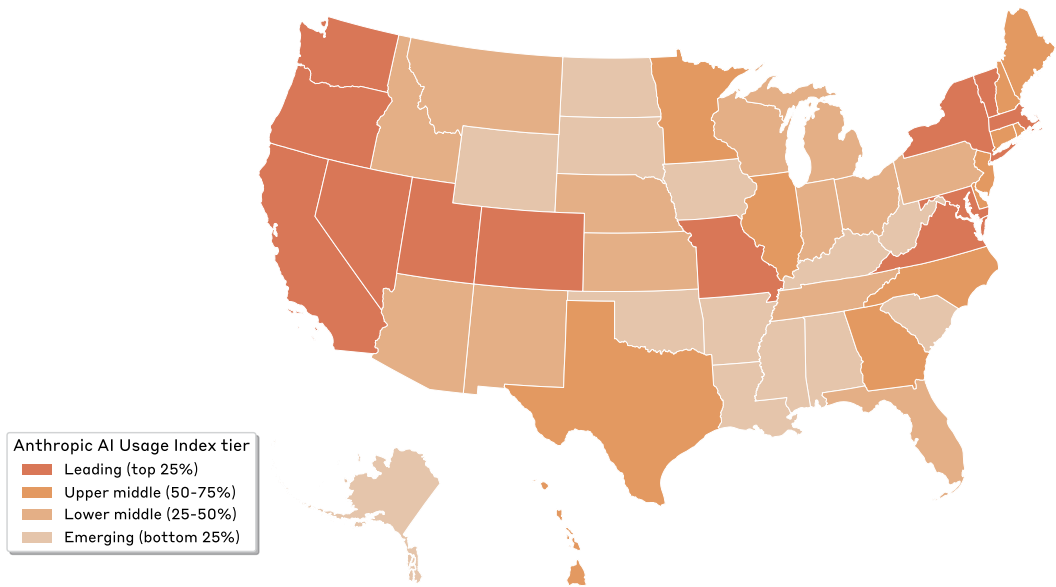

**Figure 2.6: Claude usage varies across US states, with high per-capita usage in the West Coast, but also higher usage in Nevada, Utah, Colorado, Missouri, and Virginia** The different tiers reflect a US state's position within the US distribution of the Anthropic AI Usage Index as defined in this chapter.

| Tier | AUI range | # of states | Example states |
| --- | --- | --- | --- |
| Leading (top 25%) | 0.98 - 3.82 | 13 | District of Columbia, Utah, California, New York, Virginia |
| Upper middle (50-75%) | 0.71 - 0.88 | 12 | Illinois, Connecticut, New Hampshire, New Jersey, Hawaii |
| Lower middle (25-50%) | 0.43 - 0.70 | 13 | Arizona, Florida, Pennsylvania, Tennessee, New Mexico |
| Emerging (bottom 25%) | 0.21 - 0.42 | 13 | South Carolina, Alabama, Wyoming, North Dakota, Iowa |

**Table 2.2: Claude per capita usage tiers with examples, number of states, and AUI range for each tier.**

## Task usage patterns across countries

We observe notable variation in how Claude is used in different countries. As in past reports, we analyze these trends using two different approaches. First, we classify conversations into tasks according to O*NET, a US taxonomy that maps specific tasks to occupations and occupation groups (e.g., a task involving software debugging would fall into the Computer and Mathematical occupation group).

Second, we use Claude to construct a bottom-up taxonomy of user requests on Claude.ai, which provides insight into usage patterns that do not fit neatly into existing taxonomies. For example, the request cluster "help write and improve cover letters for job applications" (lowest level) feeds into the higher-level cluster "help with job applications, resumes, and career documents" (middle level), which in turn feeds into the cluster "help with job applications, resumes, and career advancement" (highest level). These two complementary approaches allow us to both report results aligned with standard labor statistics, and provide flexibility to capture tasks that standard taxonomies miss.

**Higher per capita Claude usage is associated with more diverse task usage**

When analyzing O*NET tasks aggregated at the highest level (in terms of the Standard Occupation Classification occupation groups they belong to), we notice strong variation across countries. While the overall pattern is noisy–especially for countries with fewer observations–Figure 2.7 suggests that as we progress from lower to higher per capita Claude adoption, usage shifts away from tasks in the Computer and Mathematical occupation group (e.g., programming) to more diverse tasks in areas such as education, office and administrative uses, and arts. We also see increased usage in the life, physical and social sciences.

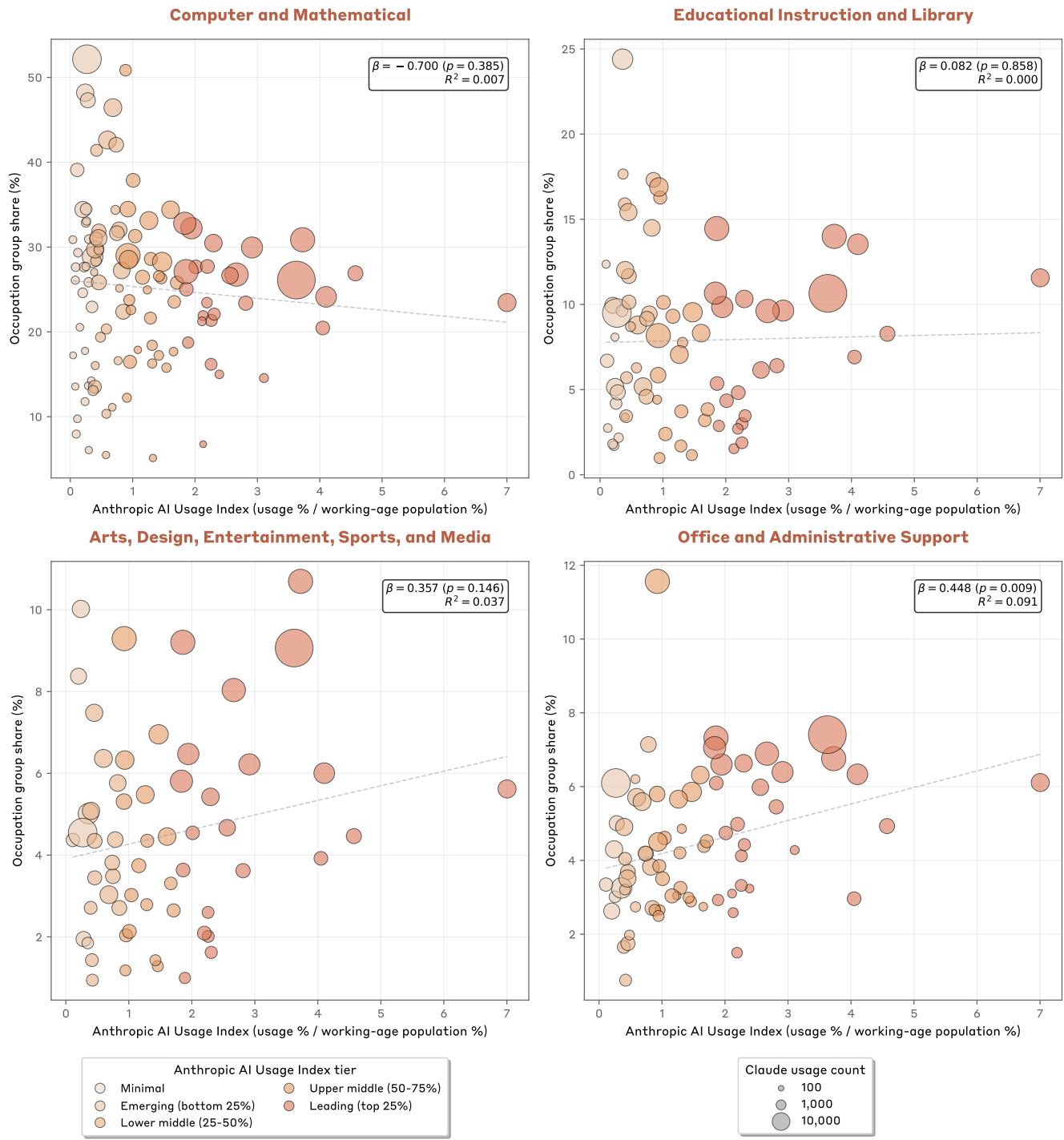

Figure 2.7: **As we move from lower to higher adoption countries, Claude usage appears to shift away from programming-dominant tasks to a more diverse mix of tasks, though the overall pattern is noisy** This figure shows the relationship between the Anthropic AI Usage Index and the most frequent Standard Occupation Classification (SOC) occupation groups. Each panel shows a different SOC group. SOC share is based on how many O*NET tasks in a given geography fall into a given SOC group. The color indicates which AUI tier a country falls into. The bubble size indicates the usage count for each country. We only include countries with at least 200 observations in our sample for this figure because of the uncertainty of the measure for low-usage countries in our random sample. The regression weights every country equally.

Country idiosyncrasies also emerge when looking at our bottom-up request taxonomy.[8] Take, for example, the United States, Brazil, Vietnam, and India, which represent the country with the highest total usage within a given Anthropic AI Usage Index tier. Users in the United States disproportionately use Claude for household management purposes, to search for jobs, and for medical guidance compared to the global average. By contrast, Claude users in Brazil have comparatively high usage for both translation and legal services. Vietnam's top disproportionate requests are related to software development and education, and India's top disproportionate requests focus almost exclusively on software development. This likely reflects local specialization: Brazil has been an [early adopter of AI in the judicial system](), and India has a large information technology sector.

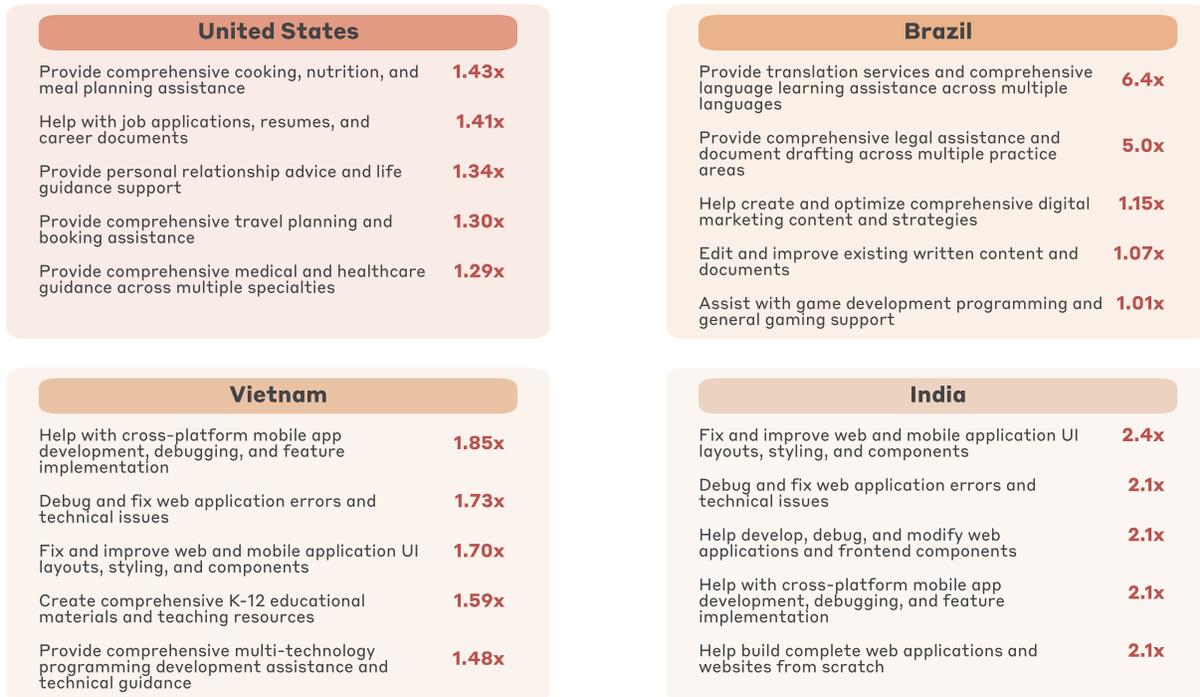

**Top overrepresented requests for the United States, Brazil, Vietnam and India**

**United States**
| Request | |
|---|---|
| Provide comprehensive cooking, nutrition, and meal planning assistance | 1.43x |
| Help with job applications, resumes, and career documents | 1.41x |
| Provide personal relationship advice and life guidance support | 1.34x |
| Provide comprehensive travel planning and booking assistance | 1.30x |
| Provide comprehensive medical and healthcare guidance across multiple specialties | 1.29x |

**Brazil**
| Request | |
|---|---|
| Provide translation services and comprehensive language learning assistance across multiple languages | 6.4x |
| Provide comprehensive legal assistance and document drafting across multiple practice areas | 5.0x |
| Help create and optimize comprehensive digital marketing content and strategies | 1.15x |
| Edit and improve existing written content and documents | 1.07x |
| Assist with game development programming and general gaming support | 1.01x |

**Vietnam**
| Request | |
|---|---|
| Help with cross-platform mobile app development, debugging, and feature implementation | 1.85x |
| Debug and fix web application errors and technical issues | 1.73x |
| Fix and improve web and mobile application UI layouts, styling, and components | 1.70x |
| Create comprehensive K-12 educational materials and teaching resources | 1.59x |
| Provide comprehensive multi-technology programming development assistance and technical guidance | 1.48x |

**India**
| Request | |
|---|---|
| Fix and improve web and mobile application UI layouts, styling, and components | 2.4x |
| Debug and fix web application errors and technical issues | 2.1x |
| Help develop, debug, and modify web applications and frontend components | 2.1x |
| Help with cross-platform mobile app development, debugging, and feature implementation | 2.1x |
| Help build complete web applications and websites from scratch | 2.1x |

**Figure 2.8: Overrepresented request clusters for the United States, Brazil, Vietnam and India** A request is overrepresented in a country when the share of conversations containing that request is higher for that country than globally. For this figure, we focus on request clusters at the middle level of granularity, i.e. more aggregated than the lowest level request clusters, but less aggregated than the highest level request clusters. Only includes requests with at least 1% frequency globally and for that country.

Across all countries, software development emerges as the most common use of Claude. Why do developer tasks consistently lead in overall Claude usage patterns? Several factors likely contribute to this effect:

- **Model-task fit:** Claude is a very strong coding model and readily deployed across code generation, debugging, and technical problem-solving tasks.
- **Developer receptivity:** Developer communities embrace new tools rapidly, and this usage diffuses through their social and professional networks.
- **Low organizational barriers:** Individual developers can typically adopt Claude without complex approval processes—in contrast to, say, medical use cases.

## Task usage patterns across the United States

In this section we explore patterns of Claude usage across states within the US, giving us further insight into how local economic conditions shape usage patterns. As we discuss above, cross-state differences in the Anthropic AI Usage Index account for less than half of the variation in income differences across US states. This suggests that other regional differences—including the compatibility of Claude capabilities with the occupational composition of the local workforce—play a larger role in determining why usage is more concentrated in some states than others.

In a number of states, we see evidence that local patterns of AI use aligns with distinctive features of the local economy. When analyzing the top states in each usage tier—California for leading, Texas for upper middle, Florida for lower middle, and South Carolina for emerging—we see strong variation in terms of our bottom-up request taxonomy (see Figure 2.9).

For example, California shows disproportionate use for IT-related requests, digital marketing and translation, likely reflecting its tech sector and linguistically diverse population. California also has disproportionately frequent requests for help with basic numerical tasks, which may represent tests of model capabilities or abuse. Florida sees disproportionate use for business advice and fitness, potentially tied to its role as a financial hub with relatively low tax rates and a warm climate amenable to outdoor activities.

**Top overrepresented high-level requests for California, Texas, Florida and South Carolina**

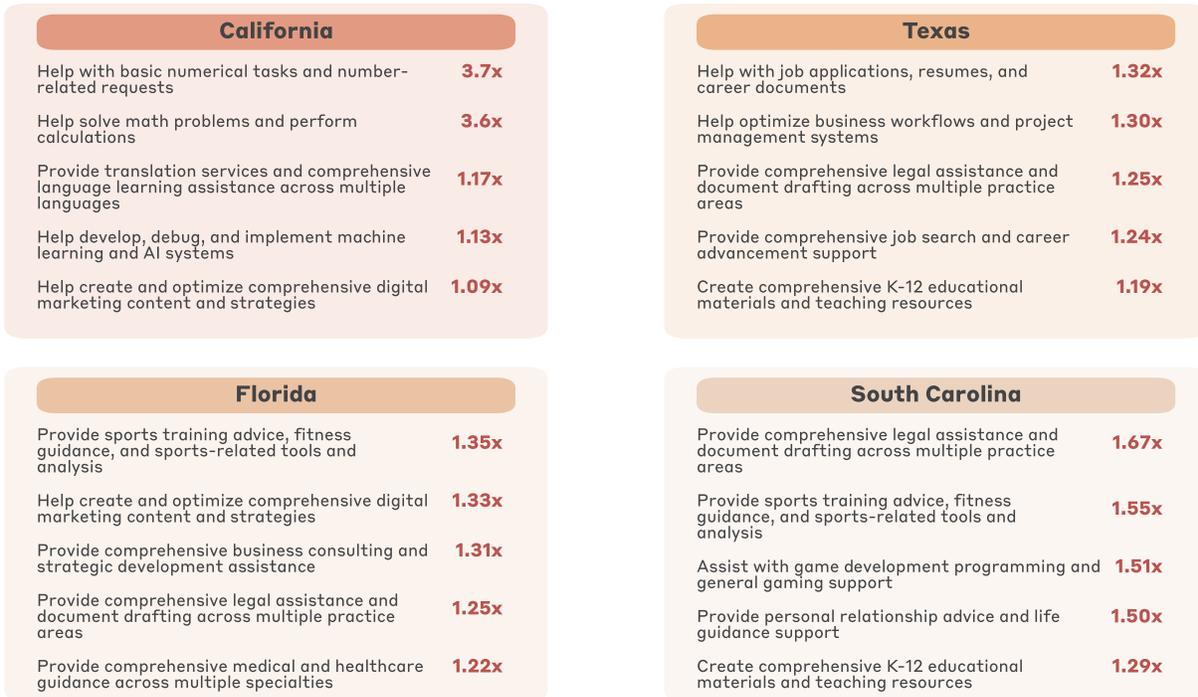

Figure 2.9: **Overrepresented request categories for California, Texas, Florida and South Carolina** A request is overrepresented in a state when the share of conversations containing that request is higher for that state than in the US as a whole. For this figure, we focus on request clusters at the middle level of granularity, i.e. more aggregated than the lowest level request clusters, but less aggregated than the highest level request clusters. Only includes requests with at least 1% frequency in the United States and for that state.

Within the US, D.C. leads in terms of per capita Claude usage, with a disproportionate focus on document editing, information provision and job applications across both the O*NET task classification and bottom-up categorization (see Figure 2.10). For example, help with job applications is 1.84x as common in DC as in the US overall. Our interactive dashboard allows everyone to explore the full range of variation and patterns across US states.

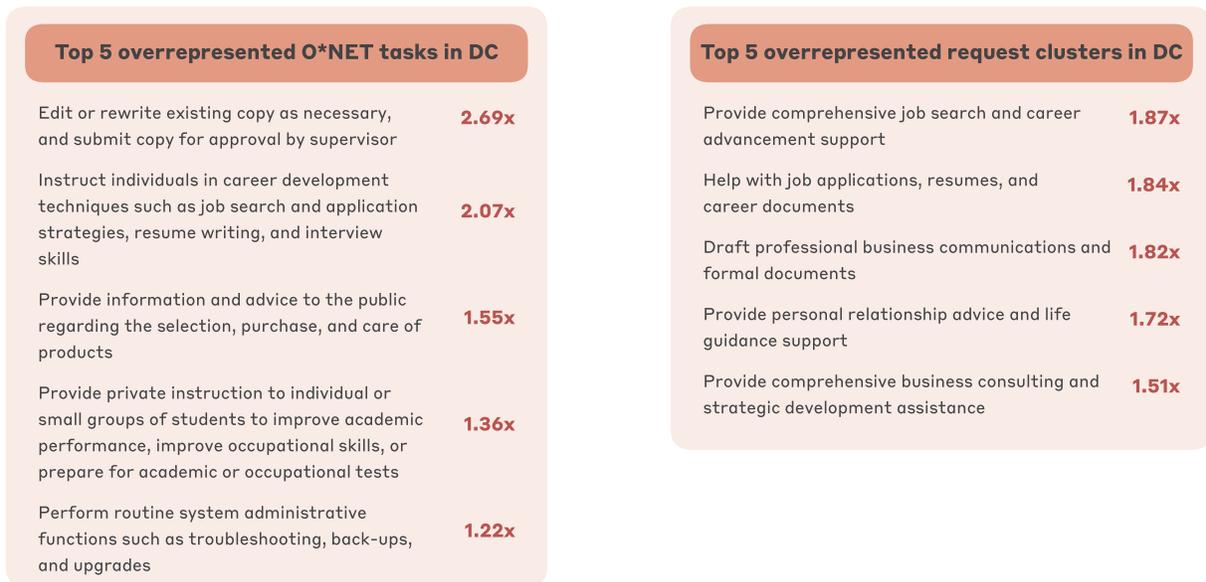

Figure 2.10: **Washington, DC has the highest Claude usage per capita, with disproportionate tasks and requests focusing on document editing, information provision and job applications** O*NET tasks refer to tasks in the O*NET taxonomy. Requests are based on the bottom-up request categories that describe what requests users make of Claude. A task or request is overrepresented in a state when the share of conversations containing that task or request is higher for that state than in the US as a whole. For this figure, we focus on request clusters at the middle level of granularity. Only includes requests with at least 1% frequency in the United States and for that state.

## Geographic patterns in human-AI collaboration

While previous sections examined what tasks people use Claude for, an equally revealing pattern emerges in how they interact with it. Here, we use the same augmentation and automation collaboration patterns as defined in Chapter 1.

Countries have different task mixes, meaning that they focus on different economic tasks, which may partly explain differences in automation patterns. In this section, we investigate whether automated use is systematically different among low and high per capita adoption economies—even when controlling for differences in task mix.[9]

We find that even when controlling for the task mix of a country, users from different countries show notably different preferences for autonomous delegation versus collaborative interaction. As Claude usage per capita increases, countries shift from automation-focused to augmentation-focused usage. This is somewhat counter-intuitive, since we are controlling for the more diverse task composition across different countries. We speculate that cultural and economic factors might affect the automation share, or perhaps that early adopters in each country tend to use AI in a more automotive way—but more research is needed here.

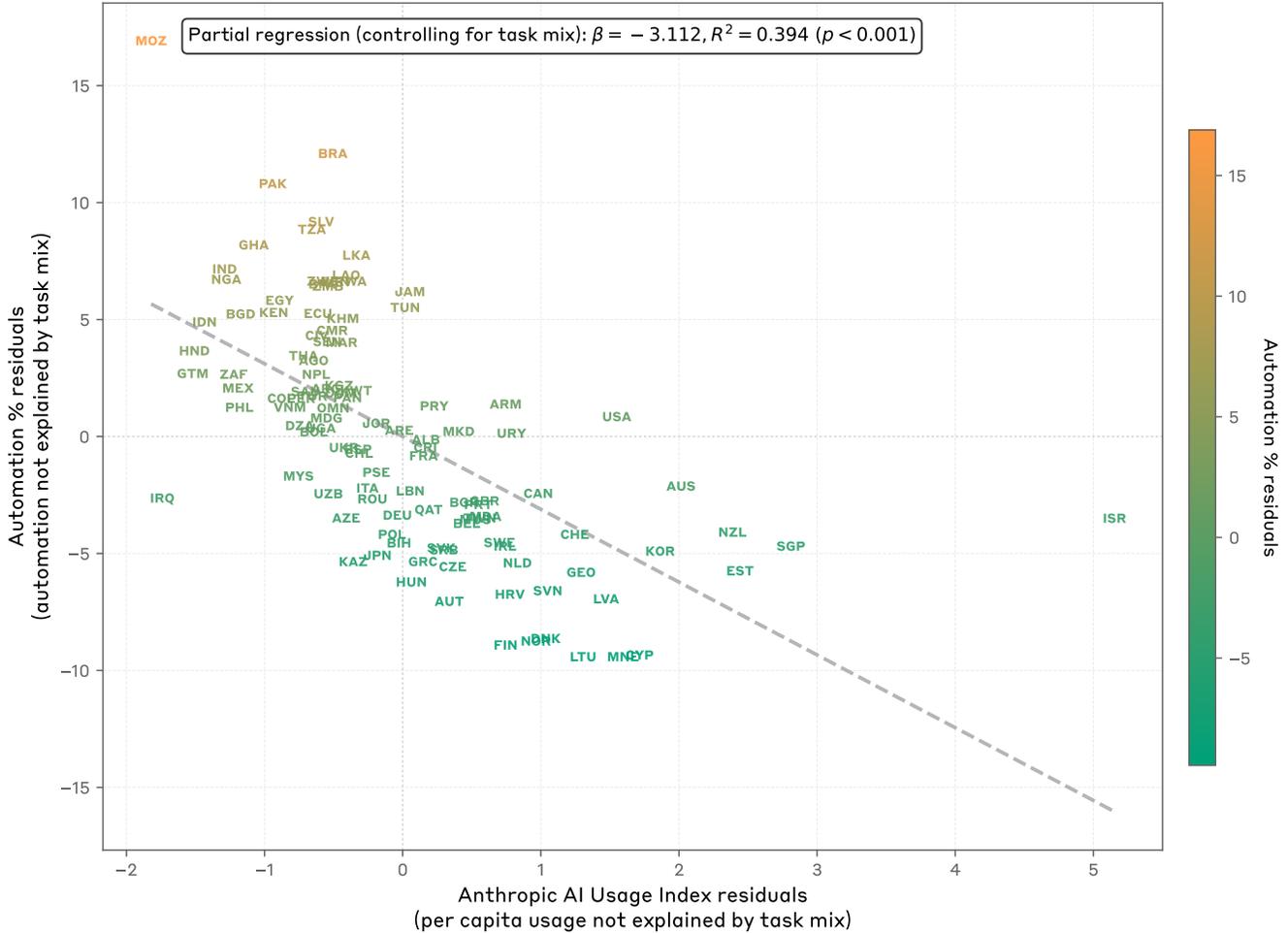

Figure 2.11: **Countries with higher Anthropic AI Usage Index tend to use Claude in a more collaborative manner (augmentation), rather than have it operate independently (automation)** This figure shows the relationship between the Anthropic AI Usage Index and the automation share in a given country. We plot the relationship after accounting for a geography's task mix, thus we show the regression residuals. We only include countries with at least 200 observations in our sample for this figure because of the uncertainty of the measure for low-usage countries in our random sample. Each country is represented by its 3-letter ISO code.

## Conclusion

Our analysis of Claude usage patterns across geographies reveals several key insights. One of the most striking is the geographic concentration of Claude usage. The leadership of the US and California in terms of Claude usage overall, and the strong correlation of Claude usage and income per capita, suggest parallels to past technologies in which initial geographic concentration and specialized use were a key feature. Drawing parallels to the diffusion patterns of prior technologies may help us better understand the diffusion and impact of AI.

Surprisingly, geography shapes not just what AI tools are used for, but how they are used. Users in economies with relatively low per capita usage have a relative preference for delegating tasks to Claude (automation), whereas users in economies with high per capita usage are somewhat more likely to prefer more collaborative or learning-based interactions with Claude (augmentation), even when controlling for the task mix. Similar to the local specialization in task use, the local specialization in AI collaboration patterns suggests that impact of AI could be very different in different regions.

The geographic patterns of AI adoption—where it is used, for which tasks, and how—suggest that in order to realize the potential of AI to benefit people across the globe, policymakers need to pay attention to local concentration of AI use and adoption, and address the risk of deepening digital divides.

---

1   For privacy reasons, our automated analysis system filters out any cells—e.g., countries, and (country, task) intersections—with fewer than 15 conversations and 5 unique user accounts. For bottom-up request clusters, we have an even higher privacy filter of at least 500 conversations and 250 unique accounts.

2   Data in this section covers 1 million Claude.ai Free and Pro conversations from August 4 to 11, 2025, randomly sampled from all conversations in that period that were not flagged as potential trust and safety violations. The unit of observation is a conversation with Claude on Claude.ai, not a user, so it is possible that multiple conversations from the same user are included, though our [past work](#) suggests that sampling conversations at random versus stratified by user does not yield substantively different results. Aggregate geographic statistics at the country and US state level were assessed and tabulated from the IP address of each conversation. For geolocation, we use ISO-3166 codes since our provider for IP geolocation uses this standard. International locations use ISO-3166-1 country codes, US state level data use ISO-3166-2 region codes, which include all 50 US states and Washington DC. We exclude conversations originating from VPN, anycast, or hosting services, as determined by our IP geolocation provider.

3   International locations use ISO-3166-1 country codes, which includes countries and some territories.

4   Tier thresholds (quartiles) are based on countries with at least 200 observations for the global level, and on US states with at least 100 observations for the US level. Countries with no observed usage are assigned to the Minimal tier since we do not know if they have exactly zero usage or little usage that our random sample did not capture. Future work, for example using stratified sampling, will allow us to explore these patterns with higher accuracy given limited observations for smaller countries and states.

5   The world map is based on Natural Earth's world map with the ISO standard point of view for disputed territories, which means that the map may not contain some disputed territories. We note that in addition to the countries shown in gray ("Claude not available"), we do not operate in the Ukrainian regions Crimea, Donetsk, Kherson, Luhansk, and Zaporizhzhia. In accordance with international sanctions and our commitment to supporting Ukraine's territorial integrity, our services are not available in areas under Russian occupation.

6   "No data" applies to countries with partially missing data. Some territories (e.g., Western Sahara, French Guiana) have their own ISO-3611 code. Some of these have some usage, others have none. Since the Anthropic AI Usage Index is calculated per working-age capita based on working age population data from the World Bank, and population data is not readily available for all of these territories, we cannot calculate the AUI for these territories.

7   When further investigating Utah's activity, we discovered a notable fraction of its usage appeared to be possibly associated with coordinated abuse. This is also reflected in a much higher "directive" automation score than average. However, we ran robustness checks and believe that this activity is not driving the results.

8   Requests were filtered to those that represent at least 1% of requests at the global level and 1% of the local level.

9   To isolate the relationship between automation preference and Claude usage accounting for task composition differences, we do the following: First, we calculate each country's expected automation percentage by taking a weighted average. For each O*NET task (e.g., coding, writing, or analysis), we multiply that task's share of the country's usage by the global automation rate for that task type (the percentage of that task that Claude completes via directive/feedback loop patterns globally). Summing these gives us the expected values for each country's automation percentage given the country's specific task mix. We then regress both the actual automation % and AUI on this expected automation %. The residuals from these regressions represent the variation in each variable that cannot be explained by task composition. By examining the relationship between these residuals (known as partial regression analysis), we can determine whether countries that have higher AI usage than their task mix would predict tend also to have higher-than-predicted automation.

**Chapter 3:**

# API Enterprise Deployment of Claude

## Overview

Whether frontier AI capabilities make us more productive, reshape labor markets, and accelerate growth will depend on when and how firms choose to deploy AI. Even when businesses recognize the potential of AI, profitably adopting it may require costly restructuring of production processes, training new workers, and other sunk-cost investments to facilitate effective deployment.[1]

To understand business adoption patterns of AI, we turn to a new data source: Anthropic's first-party (1P) API customers—again relying on [privacy-preserving methods.](#)[2] Our API allows customers to integrate Claude directly into their own products and applications, and charges by the token used, rather than a flat subscription fee. This represents a fundamentally different product experience to [Claude.ai](#), which we focused on in the previous two chapters.

Institutional inertia, alongside fixed costs of adoption, suggests that early examples of enterprise use of AI is likely to be concentrated among specialized tasks where deployment is easy, capabilities are robust, and the economic benefits from adoption are high.

Indeed, we see evidence along these lines in the data presented in this chapter. Our analysis uncovers several patterns:

- **Businesses use Claude in similar but more specialized ways than individual users.** Businesses concentrate use in tasks where AI deployment is well suited to programmatic access, like coding or administrative tasks. Compared to Claude.ai users, businesses use Claude less for educational or creative tasks and in more automated ways overall.

- **API customers tend to prefer higher cost tasks.** Despite tasks varying dramatically in cost, the most expensive tasks tend to have higher usage, suggesting that model capability, ease of deployment, and economic value of automation determine adoption much more than the cost of the interaction itself.

- **Access to appropriate contextual information is needed for sophisticated deployment.** We find evidence of an important potential bottleneck for the usefulness of AI for businesses. API customers that use Claude for complex tasks tend to provide Claude with lengthy inputs. This could represent a barrier to broader enterprise deployment for some important tasks that rely on dispersed context that is not already centralized or digitized. Correcting for this bottleneck may require firms to restructure their organization, invest in new data infrastructure, and centralize information for effective model deployment.

## Setting the stage: AI adoption patterns in public data

Before diving into our API data, it's worth grounding ourselves in the broader landscape of business AI adoption.

According to the Census Bureau's Business Trends and Outlook Survey, AI adoption among US firms has more than doubled in the past two years, rising from 3.7% in fall 2023 to 9.7% in early August 2025 (Figure 3.1).[3] Despite this rapid rate of growth, the vast majority of firms in the US do not report using AI in their production processes.

But these aggregate numbers mask large variation across sectors. For example, in early August 2025, one in four businesses in the Information sector reported using AI, which is roughly ten times the rate for Accommodation and Food Services.[4]

The picture from this public data is clear: enterprise use of AI is growing rapidly, but we are still in the early stages of AI adoption. Usage remains unevenly distributed across the economy, with the sectors most able to quickly adopt and benefit from this technology doing so.

As we will see below, our 1P API data yields a complementary conclusion: early enterprise use of Claude is likewise unevenly distributed across the economy and primarily deployed for tasks typical of Information sector occupations.

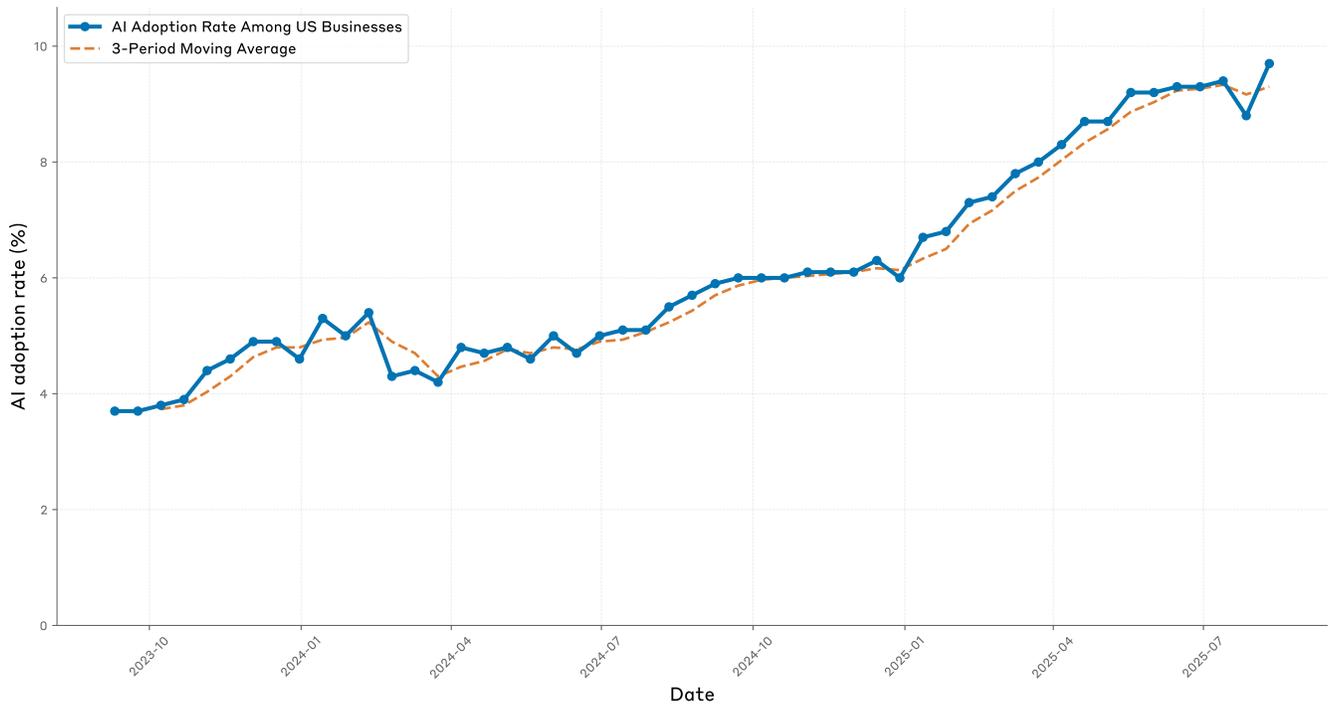

**Figure 3.1: AI adoption rates among US firms, Business Trends & Outlook Survey (Census)** Note: AI adoption rates are calculated as the share of firms responding "yes" to the question "In the last two weeks, did this business use Artificial Intelligence (AI) in producing goods or services? (Examples of AI: machine learning, natural language processing, virtual agents, voice recognition, etc.)".

## Specialized use among Anthropic API customers

To analyze API traffic, we apply the same privacy-preserving classification methods from previous chapters—categorizing anonymized API transcripts by O*NET tasks and into a bottom-up taxonomy. The patterns that emerge show enterprise usage concentrated in specialized tasks particularly suited for automation.

Overall, software development dominates the landscape. Among the top 15 use clusters—representing about half of all API traffic—the majority relate to coding and development tasks. Debugging web applications and resolving technical issues each account for roughly 6% of usage, while building professional business software represents another significant chunk. Of note, around 5% of API traffic focuses specifically on developing and evaluating AI systems themselves (Figure 3.2).

But not all API usage is for coding. API customers also deploy Claude to create marketing materials (4.7%) and to process business & recruitment data (1.9%). These two categories reveal that AI is being deployed not just for direct production of goods and services but also for talent acquisition and external communications.

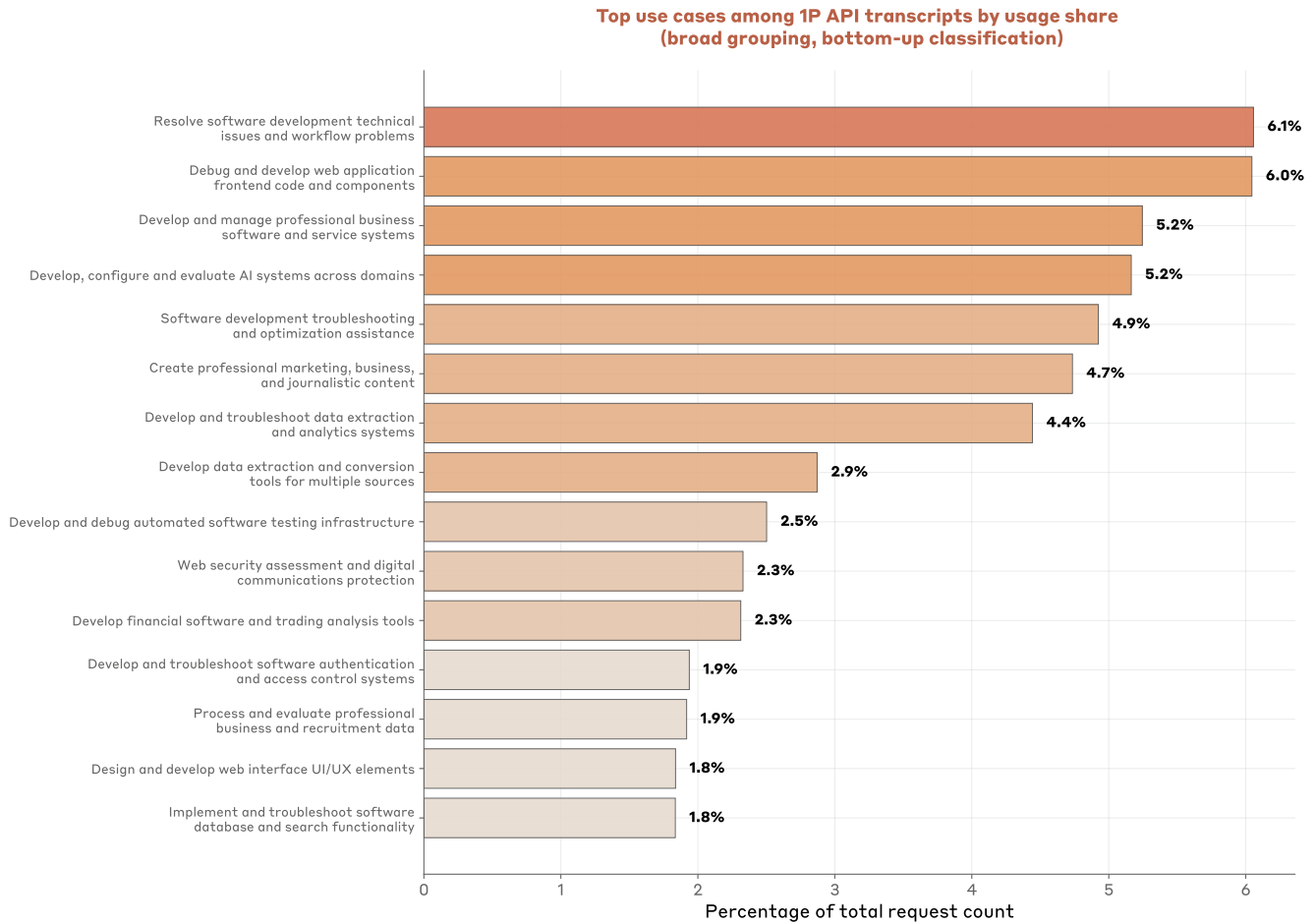

**Figure 3.2: Bottom-Up taxonomy of Claude usage among sampled 1P API transcripts** Using privacy-preserving methods we classified 1P API transcripts into a bottom-up taxonomy reflective of underlying usage. This figure reports the leading use cases at the broadest level of this taxonomy.

The O*NET classification makes these patterns even clearer. Little less than half of all API traffic maps to computer and mathematical tasks—more than 8 percentage points higher than Claude.ai usage. Office and administrative tasks come second at roughly 10% of transcripts, reflecting their suitability for automation.

On the other hand, several interaction-heavy tasks prominent on Claude.ai have a much smaller share in API usage: education and library tasks drop from 12.3% to 3.6%, while arts and entertainment fall from 8.2% to 5.2%.

In many cases however, occupational categories are reasonably close between Claude.ai and API data, suggesting that underlying model capabilities, rather than the specific product surface, drives adoption in many instances.

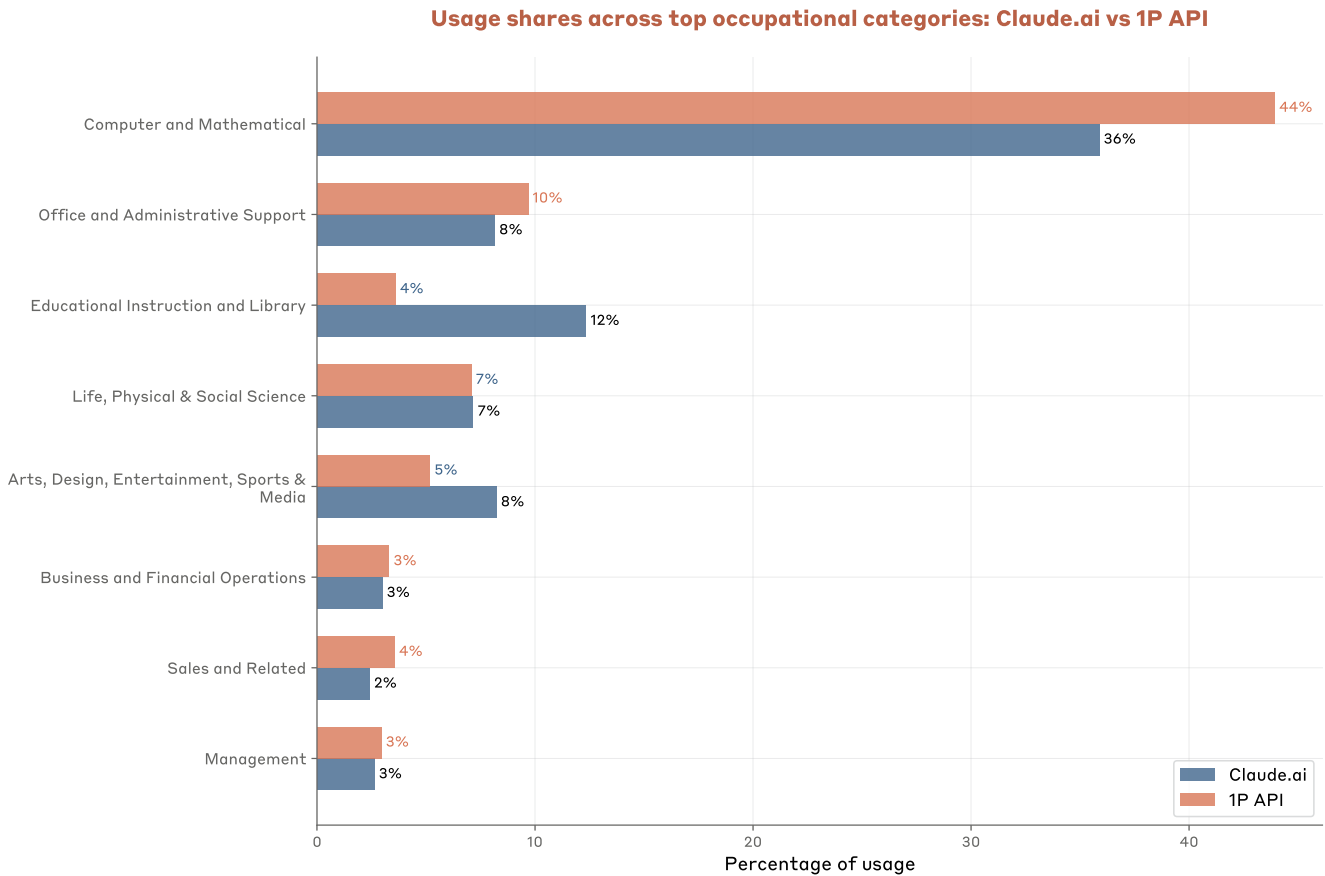

**Figure 3.3: Leading Occupational Categories by Overall Usage: Claude.ai vs 1P API** After determining usage shares for tasks, we calculate the share of traffic from Claude.ai and 1P API customers assigned to top-level occupations in the O*NET taxonomy. For example, this figure shows that 44% of API traffic in our sample was matched to a task characteristic of a Computer and Mathematical occupation.

## Occupational segmentation vs. task specialization

Despite serving different users with different interfaces, API and Claude.ai usage follows remarkably similar power law distributions across tasks. Among Claude.ai conversations, the bottom 80% of task categories account for only 12.7% of usage; for API customers it's somewhat more concentrated at 10.5% (Figure 3.4). These extreme concentrations (Gini coefficients[5] of 0.84 and 0.86) reveal massive variation in AI-task fit—the best-matched tasks see orders of magnitude more usage than poorly-matched ones.

The similarity across platforms is particularly striking given their different user bases and use cases. Both converge on comparable concentration levels, suggesting a common matching process between AI capabilities and associated economic tasks.

Tasks like code generation dominate because they hit a sweet spot where model capabilities excel, deployment barriers are minimal, and employees can adopt the new technology quickly. The long tail of rarely used tasks could reflect several factors.[6] For example, some tasks are simply less common—debugging software happens far more often than negotiating circus contracts. The extreme concentration also suggests the potential role of O-Ring[7] forces: if a task needs a level of reasoning Claude can't handle, internal data the firm can't access, or regulatory approval that doesn't exist, any single barrier could prevent adoption.

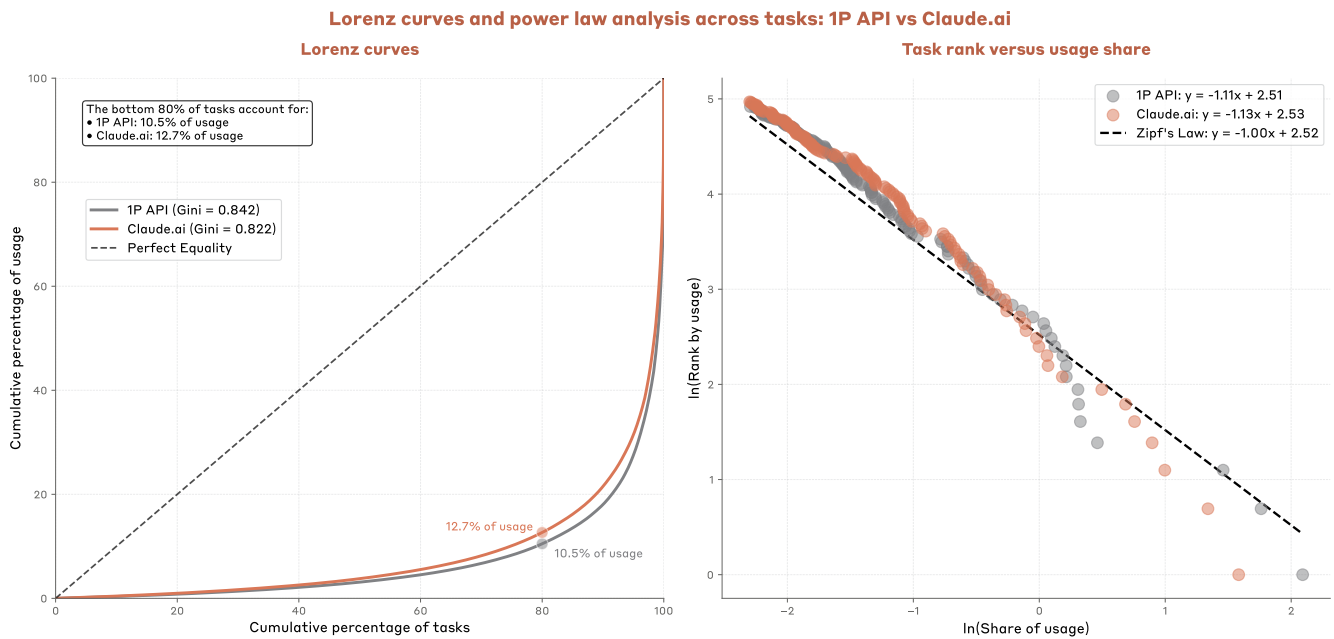

**Figure 3.4: Visualizing concentration of usage among a small number of tasks: Claude.ai versus 1P API** The left panel of this chart calculates Lorenz curves across O*NET tasks for both our Claude.ai and 1P API samples. The highlighted points on the curves indicate how much overall usage the bottom 80% of tasks account for. The right panel plots task rank against task usage share for tasks representing at least 0.1% of overall usage in our samples. Zipf's law, in which the coefficient of the best-fit-line is equal to -1, occurs with some regularity in various economic settings.

# Automation vs. augmentation among API transcripts

The clearest distinction between API and Claude.ai usage lies in how humans and AI divide the work. When businesses embed Claude into their applications, they largely delegate individual tasks rather than collaborate iteratively with models.

In our data, 77% of API transcripts show automation patterns (especially full task delegation) versus just 12% for augmentation (e.g., collaborative refinement and learning). Based on a sample of conversations from Claude.ai, the split between automation and augmentation is nearly even. Looking across economic tasks, the degree of Claude automation through the API is even starker: 97% of tasks show automation-dominant patterns in API usage, compared to only 47% on Claude.ai (Figure 3.5).

This makes intuitive sense. Programmatic API access naturally lends itself to automation: businesses provide context, Claude executes the task, and the output flows directly to end users or downstream systems.

This pattern echoes how economically consequential technologies become transformative: becoming embedded in systems that let workers access productivity gains without needing specialized skills. While both augmented and automated approaches enhance human capabilities, system-level automation is likely to yield both larger productivity gains across the economy as well as more significant changes in the labor market: Fully automating some tasks, changing which tasks are important for various jobs, and even producing new forms of work altogether.

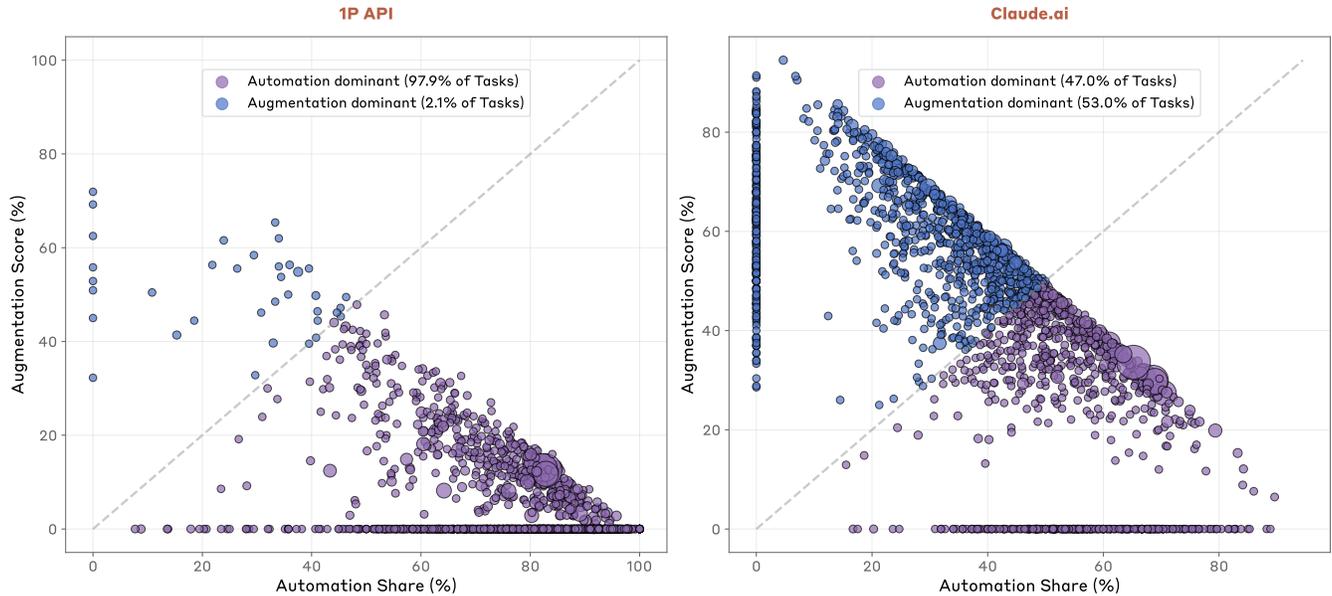

**Figure 3.5: Automation versus augmentation collaboration modes across O*NET tasks: Claude.ai versus 1P API** This figure reports the share of Claude.ai conversations and 1P API transcripts that exhibit automation or augmentation patterns of usage for each O*NET task. Automation and augmentation modes are defined in Chapter 1. When for privacy-preserving reasons we do not observe usage shares for a particular collaboration mode we give that category a value of 0% in this figure. Automation dominance is defined as a task having a greater observed share of automation usage. Likewise for augmentation dominance.

## The more Claude does, the more Claude needs to know

Why do our API customers use Claude for some tasks more than others? Beyond fundamental model capabilities, a potentially important explanation is that it is easier to provide Claude with the information needed for successful deployment for some tasks than others.

For example, if the goal is to have Claude refactor a module in a complex software development project, Claude may need to read—or at least explore— the entire codebase to understand which changes to make and where. For software development with centralized code repositories, access to this information is in principle straightforward.

For other tasks, the appropriate context might not be readily available, or it might be challenging to access. For example, asking Claude to develop a sales strategy for a key account might require Claude having access not only to information contained within a Customer Relationship Management system, but also to tacit knowledge located in the minds of account executives, marketers, and external contacts. All else equal, lacking access to such contextual information will make Claude less capable.

We explore this question by looking at the relationship across tasks between the average API input length (i.e., the context given to Claude) and Claude's average output length (i.e., what the model produces in response).[8]

For each O*NET task in our sample, we calculate the average input and output length of associated API transcripts. We then divide these values by the average lengths across all tasks appearing in our sample. This produces an input token index and an output token index for each task. An index value of 1.5, for example, means that the API transcripts associated with that task are 50% longer than the average across tasks.

There is considerable variation across tasks in how long Claude's API outputs are. For example, tasks at the 90th percentile of output length are more than 4x longer than tasks at the 10th percentile. Table 3.1 provides example O*NET tasks, along with a Claude Sonnet 4 summarization of the group of tasks at that part of the distribution.[9] Figure 3.6 shows that output length varies systematically across occupational categories as well.

## O*NET tasks with shorter and longer outputs

| Output token index | Example tasks | Claude summary of tasks |
|---|---|---|
| 10th percentile (value ~ 0.40) | • Answer questions regarding store merchandise<br>• Maintain website links and functionality<br>• Respond to customer complaints about services<br>• Configure email and virus protection software | Simple operational tasks requiring brief, straightforward responses. Focus on routine customer service, basic maintenance, and standard procedures with minimal complexity. |
| 50th percentile (value ~ 0.82) | • Analyze data to determine scientific signifcance and environmental correlations<br>• Examine objects for exhibit planning and display arrangements<br>• Observe and evaluate student work for progress assessment<br>• Manage projects and contribute to collaborative work | Moderate complexity tasks involving analysis, evaluation, and coordination. Balance between routine work and strategic thinking, requiring structured but detailed responses. |
| 90th percentile (value ~ 1.75) | • Develop new biological research methods<br>• Create experimental designs and analytical methods<br>• Revise business plans for online business<br>• Study technical blueprints and specifications | Complex analytical and development tasks requiring detailed, comprehensive responses. Focus on research, strategic planning, technical design, and creative problem-solving. |

**Table 3.1: Example O*NET tasks with shorter and longer output lengths with Claude's summaries** For each O*NET task

matched to 1P API traffic we calculate an output token index: Dividing the average output length across transcripts associated with that task by the average (unweighted) value across all tasks in our sample. Claude was prompted to identify tasks at the 10th, 50th, and 90th percentile of the output token index distribution with the minimal guidance: "The columns should be 'Example tasks', 'Index Value', 'Summary' where you provide a summary". Claude associated output length with task complexity.

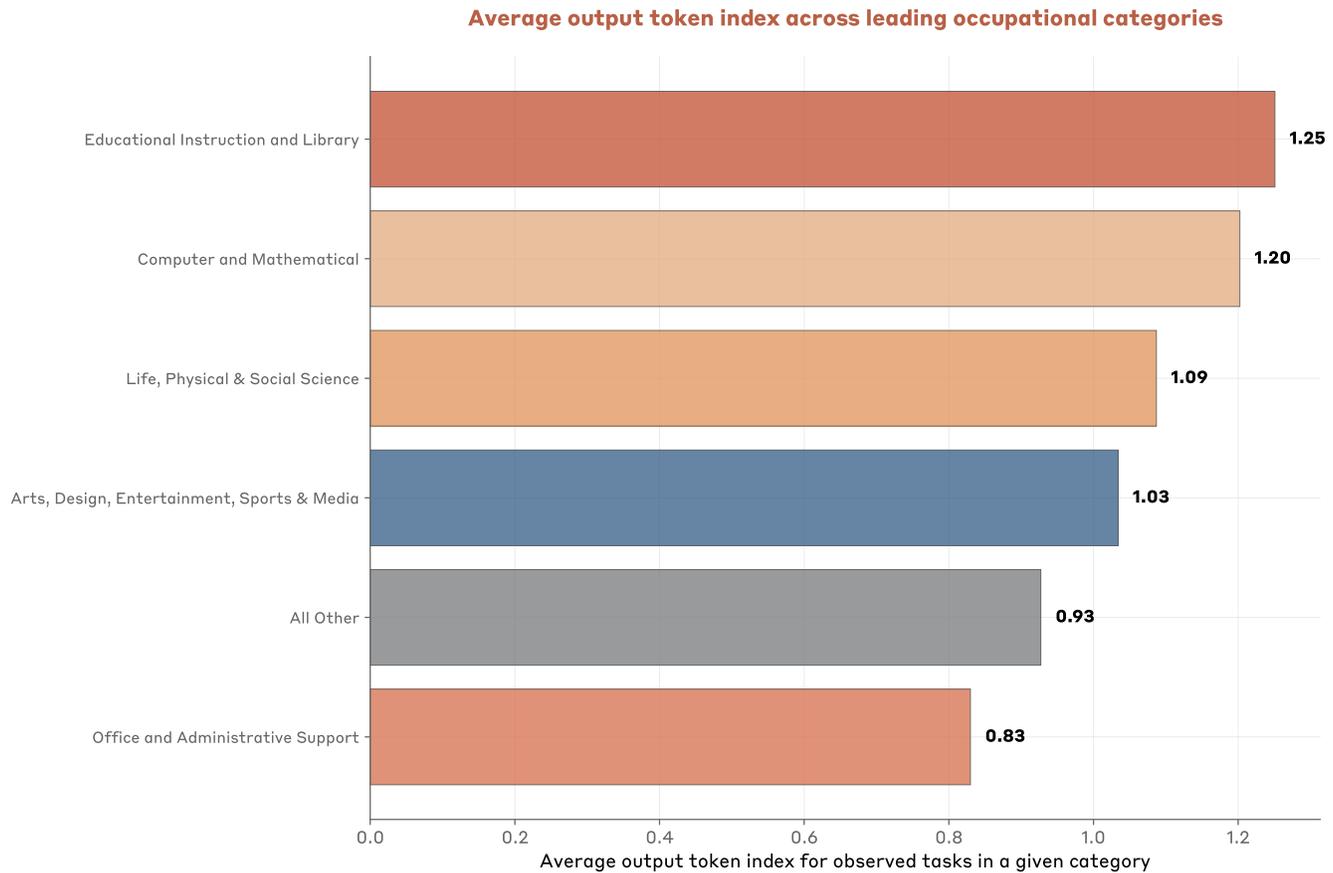

**Figure 3.6: Average output token index across O*NET tasks among leading occupational categories** For each O*NET task matched to 1P API traffic we calculate an output token index: Dividing the average output length across transcripts associated with that task by the average (unweighted) value across all tasks in our sample. We then average across tasks for a given top-level occupational categories in the O*NET taxonomy for top use occupational groups. 'All Other' combines remaining occupational groups into a single category.

What stands out from Claude's assessment of tasks is that longer output tasks tend to represent increasingly complex uses. Of course, output length does not capture all dimensions of task complexity, but it appears to be a sensible, easily measured proxy.

Because API customers are priced on the margin for both input tokens and output tokens, they have an incentive to optimize model prompting to minimize both input and output tokens when using Claude. In turn, any systematic relationship between input length and output produced by Claude partly captures the underlying contextual constraints in deploying Claude for sophisticated tasks. Stated differently, API customers are incentivized to only provide Claude with just enough context to accomplish their objective and no more. And so we learn about contextual requirements for tasks with varying output length.

Looking across tasks, we see a very stable relationship between how much context API customers provide to Claude and how much Claude actually produces. Across economic tasks, each 1% increase in input length is associated with a less-than-proportional 0.38% increase in output length (Figure 3.7). This elasticity of 0.38 suggests that there are strong diminishing marginal returns in translating longer contextual inputs into longer outputs for these economically useful tasks.[10]

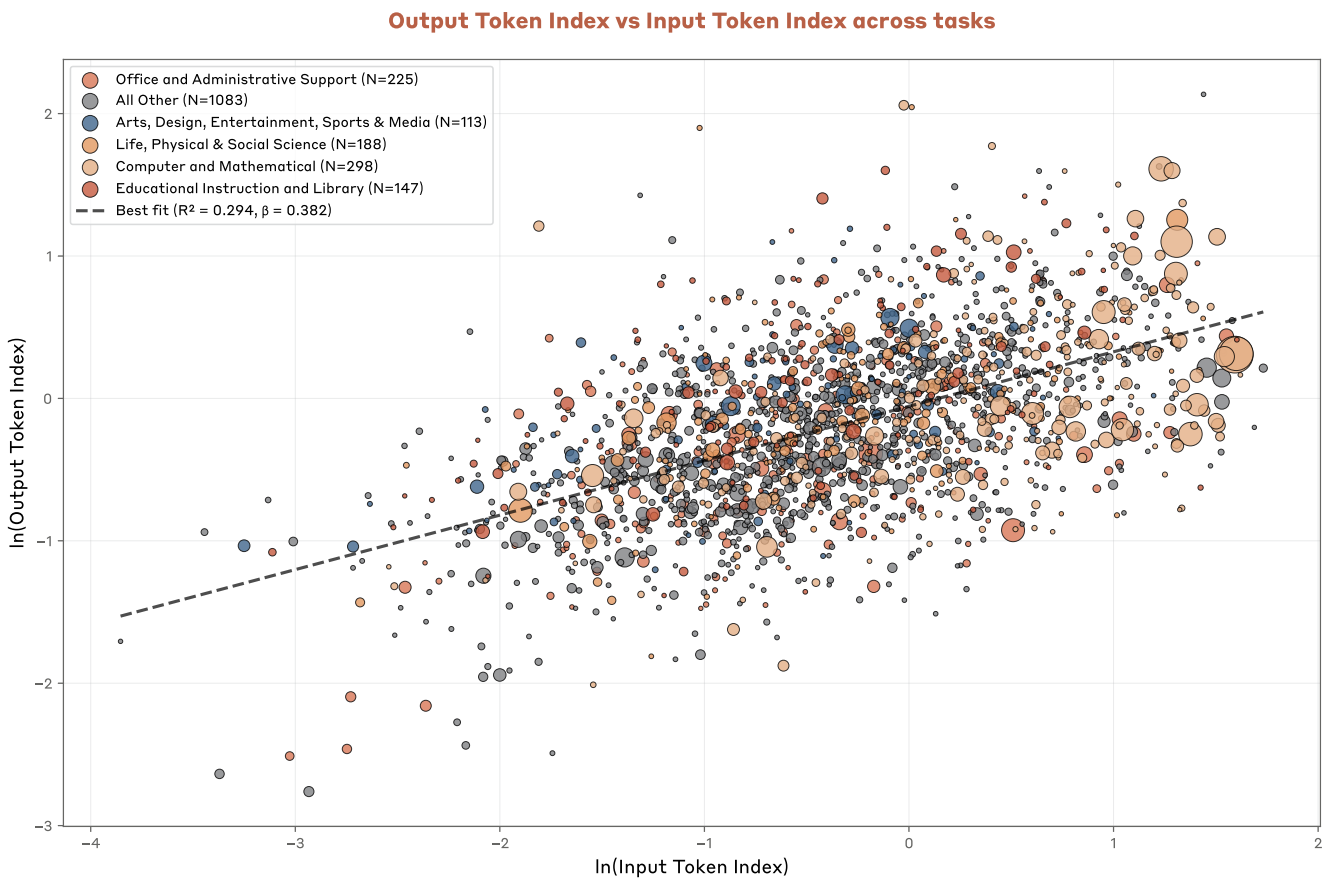

**Figure 3.7: Scatter plot of output token index and input token index across O*NET Tasks** For each O*NET task matched to 1P API traffic we calculate an output token index: Dividing the average output length across transcripts associated with that task by the average (unweighted) value across all tasks in our sample. The input token index is constructed similarly. The elasticity of 0.38 implies that each 1% increase in the input token index is associated with a 0.38% increase in the output token index.

The upshot is that deploying AI for complex tasks might be constrained more by access to information than on underlying model capabilities. Companies that can't effectively gather and organize contextual data may struggle with sophisticated AI deployment, creating a potential bottleneck for broader enterprise adoption—particularly for occupations and in industries where tacit, diffuse knowledge is crucial to business operations.

## Cost per task and substitution patterns across tasks

API customers pay per token, creating variation in the cost of deploying Claude for different tasks. More sophisticated tasks will tend to cost more, given their higher input and output token counts. This variation helps us explore whether cost is a major factor in determining which tasks businesses choose to automate with Claude.

The data suggests it is not, at least relatively speaking.[11] For example, tasks typical of computer and mathematical occupations cost more than 50% more than sales-related tasks, yet dominate usage.[12] Overall, we find a positive correlation between cost and usage: higher-cost tasks tend to have higher usage rates (Figure 3.8).

The positive correlation between cost and usage suggests that cost plays an immaterial role in shaping patterns of enterprise AI deployment. Instead, businesses likely prioritize use in domains where model capabilities are strong and where Claude-powered automation generates enough economic value in excess of the API cost.

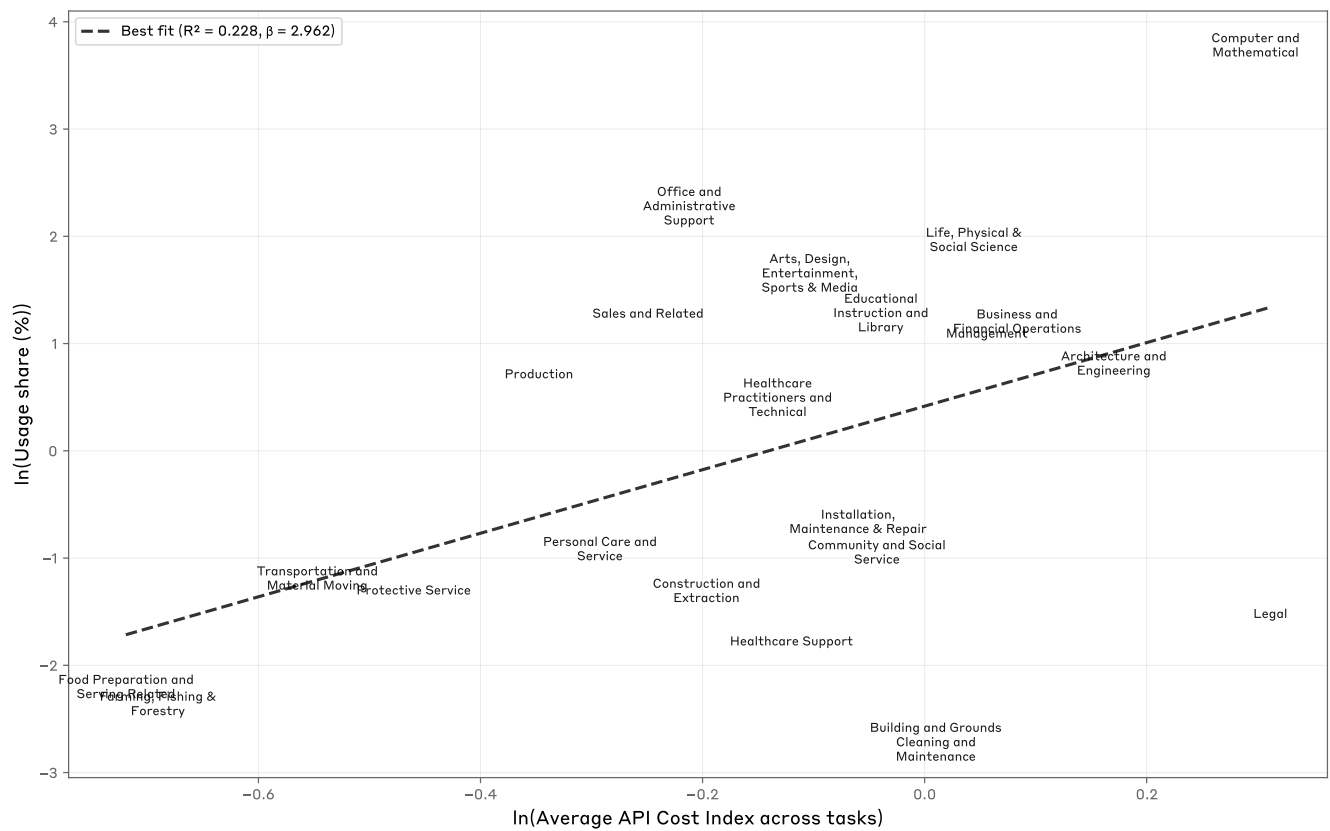

**Figure 3.8: API cost per task and usage share across occupational categories** For each O*NET task matched to 1P API traffic we calculate an API cost index: Dividing the average API cost across transcripts associated with that task by the average (unweighted) value across all tasks in our sample. This figure plots the average API cost index across tasks in a given occupational category against usage share. The estimated elasticity of 3 implies that each 1% increase in the average cost of a task is associated with a 3% increase in prevalence in our sample.

While this positive correlation holds overall, we next ask whether demand for Claude capabilities is lower among otherwise similar but costlier tasks. With the important caveat that this should be viewed as a preliminary exploration, this is what we find.

Controlling for task characteristics, we find that each 1% cost increase is associated with a 0.29% reduction in usage frequency in our sample of API transcripts (Figure 3.9).[13] While consistent with standard economic theory that higher prices lead to lower demand, the implied increase in usage to a drop in cost is limited. According to this estimate, a 10% cost reduction for a particular task would only increase usage by around 3%.

Other factors, beyond the cost of using Claude for particular tasks, appear to matter more for patterns of use.

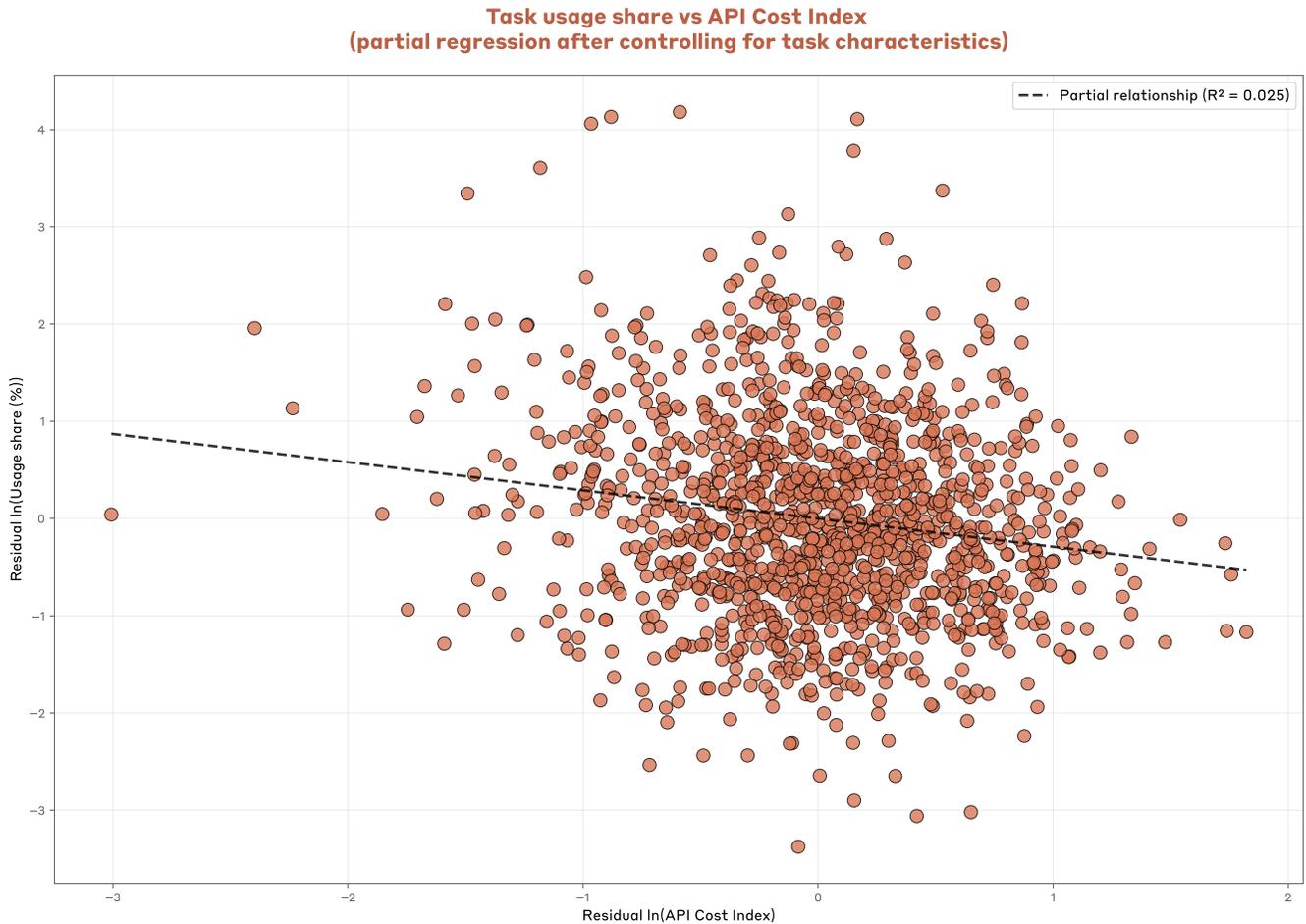

**Figure 3.9: Scatter plot of API cost per task and usage share controlling for task characteristics** For each O*NET task matched to 1P API traffic we calculate an API cost index: Dividing the average API cost across transcripts associated with that task by the average (unweighted) value across all tasks in our sample. We then restrict the sample to tasks appearing in both our 1P API and Claude.ai samples. This partial scatter plot controls for the following task-level characteristics: fixed effects for occupational category, collaboration mode share from Claude.ai, and indicators for whether a given collaboration mode was censored for privacy-preserving reasons in the Claude.ai sample. The estimated elasticity of -0.29 implies that each 1% increase in the API cost index for a given task is associated with a 0.29% decrease in prevalence in our sample, after controlling for task characteristics.

## Conclusion

Our API data captures enterprise AI adoption in its early stages: highly concentrated, automation-focused, and surprisingly price-insensitive (at least among the tasks our API customers use Claude for).

The 77% automation rate suggests enterprises use Claude to delegate tasks, rather than as a collaborative tool. Such systematic deployment is likely to be an important conduit by which AI delivers broader productivity gains within the economy. Given clear automation patterns in business deployment, this may also bring disruption in labor markets, potentially displacing those workers whose roles are most likely to face automation.

But the implications for the labor market are not entirely clear. As we document above, complex tasks require disproportionately more context. Such information may be scattered across organizations. In such conditions, workers with tacit knowledge about business operations may stand to benefit as complements to sophisticated AI-powered automation.[14] Understanding the uneven labor market implications of AI adoption is an important area for future research.

Businesses looking to adopt AI effectively may need to restructure how they organize and maintain the information that frontier systems rely on. Whether today's narrow, automation-heavy adoption evolves toward broader deployment will likely determine AI's future economic impacts.

---

[1] In the presence of fixed costs of adjustment, the question businesses face is not necessarily if they will adopt AI, but when. See Hall and Kahn 2003, *Adoption of New Technology*: "The most important thing to observe about this kind of decision is that at any point in time the choice being made is not a choice between adopting and not adopting but a choice between adopting now or deferring the decision until later."

[2] Data in this section covers 1 million transcripts from August 2025, sampled randomly from a pool of 1P API customers constituting roughly half of our 1P API usage. We continue to manage data according to our privacy and retention policies, and our analysis is consistent with our terms, policies, and contractual agreements. Each record is a prompt-response pair from our sample period which in some instances is mid-session for multi-turn interactions.

[3] Note that this is a different measure of adoption than in the introduction to this report. Reported adoption by consumers and employees of AI reached 40% in 2024 whereas when measured at the firm-level, nine out of ten businesses in the US report not using AI.

[4] The Business Trends and Outlook Survey (BTOS), published by the Census Bureau, is a reputable barometer of AI adoption by firms in the US. The survey question we use to measure AI adoption is "In the last two weeks, did this business use Artificial Intelligence (AI) in producing goods or services? (Examples of AI: machine learning, natural language processing, virtual agents, voice recognition, etc.)". See Crane, Green, and Soto 2025, *Measuring AI Uptake in the Workplace* for a comparison of BTOS with other measures of overall AI adoption among firms.

[5] The Gini coefficient is a measure used to quantify inequality within a distribution, such as the distribution of task usage. It ranges from 0 to 1, where 0 represents perfect equality (every task has exactly the same usage share) and 1 represents perfect inequality (where one task accounts for all usage, and every other task has none).

[6] Power laws in economic settings are an empirical regularity with notable examples of Zipf's law in particular. Models that generate this type of outcome feature both underlying heterogeneity and intentional, optimizing decision-making. For more, see Gabaix 2016, *Power Laws in Economics: An Introduction.*

[7] Kremer 1993, *The O-Ring Theory of Economic Development.*

[8] API input length refers to the text in API messages, system prompts, and any additional content sent to the model, including files and datasets relevant to the task at hand. Output length refers to Claude's generated response to an API call.

[9] Claude was prompted to identify tasks at the 10th, 50th, and 90th percentile of the ONET task distribution with the

minimal organization of "The columns should be 'Example tasks', 'Index Value', 'Summary' where you provide a summary".

[10] Another contributing factor could be the degradation in performance some models experience at longer context lengths. See Liu et al, 2023, *Lost in the Middle: How Language Models Use Long Contexts.*

[11] The question we ask in this section is whether, all else equal, cost differences across tasks shapes relative usage patterns. This is different from studying whether overall Claude usage is sensitive to external competitive pricing pressures.

[12] To see that this is the case, we first aggregate O*NET tasks that we identify in our API sample by broad occupational category to measure overall usage shares and the average cost per task in each category. As with the input and output tokens reported by O*NET task, we normalize average cost per task by the average value across tasks observed in our sample.

[13] Controls include fixed effects for broad occupational categories as well as collaboration mode shares by task from our concurrently sampled Claude.ai conversations. Because some tasks have censored collaboration mode shares, we also include indicators for whether that a particular mode has missing data. We restrict attention to the set of tasks identified in both our API sample and our Claude.ai samples.

[14] For example, see Ide and Talamaś, 2025, *Artificial Intelligence in the Knowledge Economy.*

# Concluding remarks

This third iteration of the Anthropic Economic Index Report captures AI adoption at a critical juncture. Existing capabilities of Claude and other frontier AI systems are already poised to transform economic activity, given how broadly applicable the technology is. Rapidly advancing AI capabilities only reinforce the conclusion that immense change is on the horizon.

And yet early AI adoption is strikingly uneven. Usage currently clusters in a small set of tasks, with strong geographic variation that is highly correlated with income—particularly across countries. Such concentration reflects where AI capabilities, ease of deployment, and economic value align: coding and data analysis have high usage, while tasks requiring dispersed context or complex regulatory navigation are further behind.

Early business adoption of Claude is at once both similar to consumer use (coding is the most common use for both), and different in several consequential ways. In particular, with programmatic access to Claude through the API, businesses tend to use Claude with greater automation. Such systematic enterprise deployment reflects how AI is poised to reshape economic activity: increasing overall productivity, but with uncertain implications for those workers whose existing responsibilities have been automated.

These patterns risk creating divergence. If AI's productivity gains concentrate in already-prosperous regions and automation-ready sectors, existing inequalities could widen rather than narrow. If AI automation improves the productivity of workers with tacit organizational knowledge—as some of our evidence suggests—then more experienced workers could see rising demand and higher wages even as entry-level workers face worse labor market prospects.[1]

Building on our previous releases, this iteration of the Index's reports marks a significant expansion in both scope and transparency. We are now open-sourcing comprehensive API usage data alongside our existing Claude.ai consumer data (now including geographic breakdowns at state and country levels), all intersected with detailed task-level classifications.

By making this data public, we hope to enable others to investigate questions we haven't considered, test hypotheses about AI's economic impacts, and develop policy responses grounded in empirical evidence.

Ultimately, the economic effects of transformative AI will be shaped as much by technical capabilities as by the policy choices societies make.

History shows that the patterns of technological adoption aren't fixed: they shift as the technologies mature, as complementary innovations emerge, and as societies make deliberate choices about their deployment. The patterns of highly concentrated use that we observe today may yet evolve towards a broader distribution—one that captures more of AI's productivity-enhancing potential, accelerates innovation in lagging sectors, and enables new forms of economic value creation.

We are still in the early stages of this AI-driven economic transformation. The actions that policymakers, business leaders and the public take now will shape the years to come. We'll continue tracking these patterns as AI capabilities advance, and provide empirical grounding for navigating one of the most significant economic transitions of our time.

---

[1] Brynjolfsson, Chandar, and Chen 2025, *Canaries in the Coal Mine? Six Facts about the Recent Employment Effects of Artificial Intelligence* documents clear evidence that entry-level workers with high AI exposure have had relatively worse employment prospects since late 2022. Setting aside questions of causality, the straightforward interpretation is that this is due to AI substituting for work previously done by early-career workers. An alternative interpretation is presented by Gans 2025, *If AI and workers were strong complements, what would we see?*: That relatively faster employment growth for experienced workers reflects AI making such workers more productive and thus in high demand. Whether AI compliments or substitutes work is perhaps the most important question that we hope our data will help answer.